\documentclass[twocolumn]{aastex62}
\usepackage{verbatim}
\usepackage{gensymb}
\usepackage{graphicx}
\usepackage{array}
\usepackage{booktabs}
\usepackage{threeparttable}
\usepackage{natbib}
\usepackage{bm}
\usepackage{amsmath}
\newcolumntype{+}{>{\global\let\currentrowstyle\relax}}
\newcolumntype{^}{>{\currentrowstyle}}

\begin{document}

\title{An unbiased ALMA spectral survey of the LkCa~15 and MWC~480 protoplanetary disks}

\correspondingauthor{Ryan A. Loomis}
\email{rloomis@cfa.harvard.edu}

\author[0000-0002-8932-1219]{Ryan A. Loomis}
\altaffiliation{NRAO Jansky Fellow}
\affil{National Radio Astronomy Observatory, Charlottesville, VA 22903, USA}

\author[0000-0001-8798-1347]{Karin I. {\"O}berg}
\affil{Harvard-Smithsonian Center for Astrophysics, Cambridge, MA 02138, USA}

\author[0000-0003-2253-2270]{Sean M. Andrews}
\affil{Harvard-Smithsonian Center for Astrophysics, Cambridge, MA 02138, USA}

\author[0000-0003-4179-6394]{Edwin Bergin}
\affil{University of Michigan, Ann Arbor, MI 48109, USA}

\author[0000-0002-8716-0482]{Jennifer Bergner}
\affil{Harvard-Smithsonian Center for Astrophysics, Cambridge, MA 02138, USA}

\author[0000-0003-0787-1610]{Geoffrey A. Blake}
\affil{Division of Chemistry and Chemical Engineering, California Institute of Technology, Pasadena, CA 91125, USA}
\affil{Division of Geological and Planetary Sciences, California Institute of Technology, Pasadena, CA 91125, USA}

\author[0000-0003-2076-8001]{L. Ilsedore Cleeves}
\affil{Department of Astronomy, University of Virginia, Charlottesville, VA 22903, USA}

\author[0000-0002-1483-8811]{Ian Czekala}
\altaffiliation{NASA Hubble Fellowship Program Sagan Fellow}
\affiliation{Department of Astronomy, 501 Campbell Hall, University of California, Berkeley, CA 94720-3411, USA}

\author[0000-0001-6947-6072]{Jane Huang}
\affil{Harvard-Smithsonian Center for Astrophysics, Cambridge, MA 02138, USA}

\author[0000-0003-1837-3772]{Romane Le Gal}
\affil{Harvard-Smithsonian Center for Astrophysics, Cambridge, MA 02138, USA}

\author{Francois M\'enard}
\affil{Univ. Grenoble Alpes, CNRS, IPAG (UMR 5274) , F-38000 Grenoble, France}

\author{Jamila Pegues}
\affil{Harvard-Smithsonian Center for Astrophysics, Cambridge, MA 02138, USA}

\author[0000-0001-8642-1786]{Chunhua Qi}
\affil{Harvard-Smithsonian Center for Astrophysics, Cambridge, MA 02138, USA}

\author[0000-0001-6078-786X]{Catherine Walsh}
\affiliation{School of Physics and Astronomy, University of Leeds, Leeds LS2 9JT, UK}

\author[0000-0001-5058-695X]{Jonathan P. Williams}
\affil{Institute for Astronomy, University of Hawai'i at M\={a}noa, Honolulu, HI 96822, USA}

\author[0000-0003-1526-7587]{David J. Wilner}
\affil{Harvard-Smithsonian Center for Astrophysics, Cambridge, MA 02138, USA}

\begin{abstract}
The volatile contents of protoplanetary disks both set the potential for planetary chemistry and provide valuable probes of defining disk system characteristics such as stellar mass, gas mass, ionization, and temperature structure. Current disk molecular inventories are fragmented, however, giving an incomplete picture: unbiased spectral line surveys are needed to assess the volatile content. We present here an overview of such a survey of the protoplanetary disks around the Herbig Ae star MWC~480 and the T Tauri star LkCa~15 in ALMA Band 7, spanning $\sim$36 GHz from 275--317 GHz and representing an order of magnitude increase in sensitivity over previous single-dish surveys. We detect 14 molecular species (including isotopologues), with 5 species (C$^{34}$S, $^{13}$CS, H$_{2}$CS, DNC, and C$_2$D) detected for the first time in protoplanetary disks. Significant differences are observed in the molecular inventories of MWC~480 and LkCa~15, and we discuss how these results may be interpreted in light of the different physical conditions of these two disk systems.
\end{abstract}

\section{Introduction}
    \label{SLS_S1}
    Protoplanetary disks are the formation sites of planets, and their molecular inventories regulate the composition of nascent comets and planetesimals \citep[e.g.,][]{Helling_2014}. These molecules also serve as valuable probes of disk properties such as stellar mass, temperature and density gradients, ionization, and turbulence \citep[e.g.,][]{Dutrey_2007, Oberg_2011, Rosenfeld_2012, Rosenfeld_2013, Cleeves_2015, Czekala_2015, Teague_2016, Flaherty_2018, Pinte_2018}. Disks are sufficiently cold, however, such that the a large fraction of the molecular content is locked up in ices in the disk midplane and unobservable with rotational spectroscopy. The resulting intrinsically low column densities have therefore limited the majority of disk observations thus far to targeted observations of specific molecules \citep[e.g.][]{Kastner_2018}, providing valuable insight for certain species, but still leaving incomplete inventories and a fragmented view of disk chemistry.
    
    To date, 23 molecules (35 total species including isotopologues) have been discovered in disks, with the majority detected via rotational transitions \citep{McGuire_2018}. Expanding this molecular inventory is necessary to fully assess disk chemical compositions and develop new probes of chemistry and physics.  First, many of the known species in disks have isotopologues which should be present but have not yet been detected, limiting our knowledge of processes such as nitrogen fractionation \citep[e.g.,][]{Guzman_2017} and deuterium fractionation \citep[e.g.,][]{Huang_2017}. Second, although small volatiles containing C/N/O have been well studied, S-bearing species have not yet been investigated in great detail, with studies of sulfur chemistry in disks limited mainly to CS and SO \citep[e.g.,][]{Thi_2004, Dutrey_2007, Booth_2018, LeGal_2019}, and a recent detection of H$_2$S \citep{Phuong_2018}. Thus there are a number of potentially abundant S-bearing species that simply have not been searched for with deep integration times. Third, the recent detections of the complex organic molecules (COMs) CH$_3$CN, CH$_3$OH, and HCOOH in disks \citep{Oberg_2015, Walsh_2016, Bergner_2018, Loomis_2018_CH3CN, Favre_2018} suggest that more complex species may be present. Indeed, disk chemistry models predict that a significant fraction of volatiles may be in the form of COMs as they form readily in irradiated ices \citep[e.g.,][]{Bennett_2007, Garrod_2008, Oberg_2009}, and the ramifications of the resultant chemical CO depletion are extremely important to characterize when considering disk masses \citep[e.g.,][]{Miotello_2016, Miotello_2017, Yu_2017} and C/O ratios \citep[e.g.][]{Schwarz_2018}. 

    Unbiased spectral line surveys offer the potential to fill these gaps. Historically used as a powerful tool to probe the molecular inventories of cold clouds and star-forming regions \citep[e.g.,][]{Johansson_1984, Blake_1986, Blake_1994, vanDishoeck_1995, Kaifu_2004, Belloche_2008a, Remijan_2009}, single-dish surveys have also expanded our understanding of molecular complexity through a large number of serendipitous molecular detections \citep[e.g.,][]{Belloche_2008b, Belloche_2009, Ossenkopf_2010, Pety_2012, McGuire_2012, Loomis_2013, Zaleski_2013, McGuire_2016}. Recent advances in sensitivity have additionally allowed similar single-dish survey exploration of disks \citep[e.g.][]{Kastner_2014, Punzi_2015}. With the construction of the Atacama Large Millimeter/submillimeter Array (ALMA), these sensitive line surveys can now be conducted efficiently ($\sim$20 min. on-source integration per source per setting) with simultaneous high resolution imaging \citep[e.g.][]{Jorgensen_2016, Belloche_2017}, providing a clear opportunity for unbiased interferometric line surveys of disks.
    
    We have undertaken a spectral line survey of the protoplanetary disks around MWC~480 and LkCa~15 in ALMA Band 7, spanning $\sim$36 GHz from 275--317 GHz and representing an order of magnitude improvement in sensitivity compared to previous single dish surveys of disks \citep[e.g.][]{Kastner_2014, Punzi_2015}. Both MWC~480, a Herbig Ae star, and LkCa~15, a T Tauri star, reside in the nearby Taurus star forming region \citep[$\sim$160pc][]{GAIA}. They are both relatively young ($\sim$3--7 Myr) and host large ($>$200 AU) well-studied gas-rich disks \citep[e.g.,][]{Chiang_2001, Pietu_2007, Oberg_2011, Isella_2012, Huang_2017}. Differences in the luminosity, disk mass, and temperature between these two sources (see Table \ref{Table_sources}) allow for a preliminary investigation of the effect these parameters play in setting the disk molecular inventory. 
    
    \begin{deluxetable*}{cccccccccccc}
    \tablecaption{Disk properties\label{Table_sources}}
    \tablecolumns{12}
    \tablewidth{\columnwidth}
    \tablehead{
    	\colhead{Source} & \colhead{Stellar Type} &  \colhead{R.A.$^{a}$}  & \colhead{Dec.$^{a}$} & \colhead{Dist.$^{a}$} &  \colhead{L$_{\star}$} &  \colhead{M$_{\star}$} & \colhead{Disk Mass} & \colhead{Incl.} & \colhead{P.A.} & \colhead{Age} & \colhead{V$_{\mathrm{LSRK}}$} \\ [-1ex]
                &               &  \colhead{(J2000)} & \colhead{(J2000)} & \colhead{(pc)} & \colhead{(L$_{\sun}$)} & \colhead{(M$_{\sun}$)} & \colhead{(M$_{\sun}$)} & \colhead{(deg)} & \colhead{(deg)} & \colhead{(Myr)} & \colhead{(km/s)}}  
        \startdata
        MWC~480 & A1-A3/4$^{[1,2]}$   & 04:58:46.3    & 29:50:37.0 & 162 & 19-24$^{[3,4]}$  & 1.7-2.3$^{[3,4,5]}$       & 0.11$^{[6]}$      & 37$^{[7]}$    & 148$^{[7]}$ & 6-7.1$^{[3,4,5]}$ & 5.1$^{[7]}$ \\
        LkCa~15 & K3-K5$^{[2,8]}$   & 04:39:17.8    & 22:21:03.4  & 159 & 0.8$^{[3]}$     & 1.0$^{[3,5]}$           & 0.05-0.1$^{[9]}$   & 52$^{[7]}$    & 60$^{[7]}$ & 3-5$^{[3,5,10]}$ & 6.3$^{[7]}$ \\
        \enddata
    
    \tablenotetext{}{$^{a}$ Right ascension, declination, and distance of each source are from the Gaia DR2 catalog \citep{GAIA}.}
    \tablenotetext{}{\textbf{References}---[1] \cite{The_1994}, [2] \cite{Luhman_2010}, [3] \cite{Andrews_2013}, [4] \cite{Mannings_1997}, [5] \cite{Simon_2000}, [6] \cite{Chiang_2001}, [7] \cite{Huang_2017}, [8] \cite{Herbig_1988}, [9] \cite{Isella_2012}, [10] \cite{Guilloteau_2014}.}
    \end{deluxetable*}
    
    In this paper, we present an overview of the line survey and summarize the molecular content of both disks. We find evidence for 14 molecular species in total (including isotopologues), with 5 species detected in a protoplanetary disk for the first time. We present the details of the observations and the data calibration in \S\ref{SLS_S2}. In \S\ref{SLS_S3}, we describe our data analysis methods, in which the survey is imaged in an unbiased manner and a matched filtering technique is used to efficiently identify lines. \S\ref{SLS_S4} then presents an overview of the imaged data, as well as images and spectra for each molecular detection. In \S\ref{SLS_S5}, we compare the molecular inventory of the two sources and discuss how their physical characteristics may relate to the observed chemical differences. We additionally briefly discuss the low degree of chemical complexity observed in the context of predictions from chemical models. A summary is given in \S\ref{SLS_S6}.

\section{Observations}
\label{SLS_S2}
    \begin{deluxetable*}{cccccccc}
    \tablecaption{Observation details\label{SLS_Table1}}
    \tablecolumns{8}
    \tablewidth{\columnwidth}
    \tablehead{                                                                           
        \colhead{Setting$^a$} &   \colhead{Date}  &   \colhead{Antennas$^b$}&   \colhead{Baselines (m)}   &   \colhead{On-source int. (min)$^c$}&   \colhead{Bandpass Cal.}   &   \colhead{Phase Cal.}  &   \colhead{Flux Cal.}}

        \startdata
        A       &   2016 Jan. 17    &   36          &   15--331         &   19.2                    &   J0510+1800      &   J0438+3004  &   J0510+1800 \\
        B       &   2016 Jan. 17    &   31          &   15--331         &   17.6                    &   J0510+1800      &   J0438+3004  &   J0510+1800 \\
                &   2016 Apr. 23    &   36          &   15--463         &   12.6                    &   J0238+1636      &   J0433+2905  &   J0510+1800 \\
                &   2016 Dec. 12    &   36          &   15--650         &   20.7                    &   J0510+1800      &   J0438+3004  &   J0510+1800 \\
        C       &   2016 Dec. 13    &   39          &   15--650         &   20.2                    &   J0510+1800      &   J0438+3004  &   J0510+1800 \\
        D       &   2016 Dec. 13    &   36          &   15--650         &   12.6                    &   J0510+1800      &   J0438+3004  &   J0510+1800 \\
                &   2016 Dec. 14    &   39          &   15--460         &   12.6                    &   J0510+1800      &   J0438+3004  &   J0510+1800 \\
        E       &   2016 Dec. 17    &   40          &   15--460         &   13.1                    &   J0510+1800      &   J0438+3004  &   J0510+1800 \\
                &   2016 Dec. 18    &   42          &   15--492         &   13.1                    &   J0510+1800      &   J0438+3004  &   J0510+1800 \\
        \enddata
        
    \tablenotetext{}{$^{a}$ See Table \ref{SLS_Table3} for details on each spectral setting.}
    \tablenotetext{}{$^{b}$ Number of antennas remaining after flagging.}
    \tablenotetext{}{$^{c}$ Single source integration time. MWC~480 and LkCa~15 had equal integration times for all observations.}
    \end{deluxetable*}

    \begin{deluxetable*}{ccccccc}
    \tablecaption{Details of spectral settings\label{SLS_Table3}}
    \tablecolumns{7}
    \tablewidth{\columnwidth}
    \tablehead{                                                                      
        \colhead{Setting} &   \colhead{SPW} &   \colhead{Frequencies}     &   \multicolumn{2}{c}{Beam (PA)}           &   \multicolumn{2}{c}{Per chan. RMS (mJy bm$^{-1}$)} \\[-1ex]
                &       &   \colhead{(GHz)}           &   \colhead{MWC~480}              &   \colhead{LkCa~15}          &   \colhead{MWC~480}         &   \colhead{LkCa~15}         }
    
        \startdata 
        A       &   0   &   275.20--277.08  &   1$\farcs$30$\times$1$\farcs$05 (12.7$\degree$)      &   1$\farcs$18$\times$1$\farcs$04 (-42.9$\degree$) &   2.7             &   2.8             \\
                &   1   &   277.08--278.95  &   1$\farcs$27$\times$1$\farcs$03 (12.6$\degree$)      &   1$\farcs$15$\times$1$\farcs$02 (-43.3$\degree$) &   3.0             &   3.1             \\
                &   2   &   287.20--289.08  &   1$\farcs$22$\times$1$\farcs$00 (-9.7$\degree$)      &   1$\farcs$11$\times$0$\farcs$99 (-44.6$\degree$) &   3.3             &   3.4             \\
                &   3   &   289.08--290.95  &   1$\farcs$22$\times$1$\farcs$01 (168.9$\degree$)     &   1$\farcs$10$\times$0$\farcs$97 (-45.7$\degree$) &   3.3             &   3.4             \\
        B       &   0   &   278.95--280.83  &   1$\farcs$13$\times$0$\farcs$60 (-30.6$\degree$)     &   1$\farcs$03$\times$0$\farcs$60 (-38.4$\degree$) &   1.8             &   1.7             \\
                &   1   &   280.82--282.70  &   1$\farcs$12$\times$0$\farcs$59 (-30.4$\degree$)     &   1$\farcs$02$\times$0$\farcs$60 (-38.0$\degree$) &   1.9             &   1.8             \\
                &   2   &   290.95--292.83  &   1$\farcs$10$\times$0$\farcs$60 (-32.5$\degree$)     &   0$\farcs$98$\times$0$\farcs$58 (-38.2$\degree$) &   2.2             &   2.2             \\
                &   3   &   292.82--294.70  &   1$\farcs$07$\times$0$\farcs$57 (149.5$\degree$)     &   0$\farcs$98$\times$0$\farcs$58 (-38.3$\degree$) &   2.0             &   2.0             \\
        C       &   0   &   283.45--285.33  &   0$\farcs$99$\times$0$\farcs$53 (-177.5$\degree$)    &   0$\farcs$85$\times$0$\farcs$53 (177.3$\degree$) &   2.3             &   2.3             \\
                &   1   &   285.33--287.20  &   0$\farcs$99$\times$0$\farcs$53 (175.8$\degree$)     &   0$\farcs$85$\times$0$\farcs$53 (-2.9$\degree$)  &   2.5             &   2.6             \\
                &   2   &   295.45--297.33  &   0$\farcs$95$\times$0$\farcs$52 (-4.4$\degree$)      &   0$\farcs$83$\times$0$\farcs$52 (170.5$\degree$) &   2.6             &   2.6             \\
                &   3   &   297.33--299.20  &   0$\farcs$94$\times$0$\farcs$51 (-3.9$\degree$)      &   0$\farcs$83$\times$0$\farcs$52 (171.3$\degree$) &   2.5             &   2.5             \\
        D       &   0   &   298.30--300.18  &   1$\farcs$00$\times$0$\farcs$56 (-25.4$\degree$)     &   0$\farcs$91$\times$0$\farcs$55 (-33.8$\degree$) &   3.0             &   3.1             \\
                &   1   &   300.18--302.06  &   1$\farcs$04$\times$0$\farcs$58 (-30.0$\degree$)     &   0$\farcs$96$\times$0$\farcs$56 (141.0$\degree$) &   3.4             &   3.6             \\
                &   2   &   310.30--312.18  &   1$\farcs$00$\times$0$\farcs$56 (-29.7$\degree$)     &   0$\farcs$92$\times$0$\farcs$55 (141.2$\degree$) &   3.5             &   3.7             \\
                &   3   &   312.18--314.06  &   0$\farcs$95$\times$0$\farcs$53 (155.2$\degree$)     &   0$\farcs$86$\times$0$\farcs$53 (-33.2$\degree$) &   3.7             &   3.9             \\
        E       &   0   &   302.05--303.93  &   0$\farcs$94$\times$0$\farcs$62 (179.2$\degree$)     &   0$\farcs$82$\times$0$\farcs$63 (-8.2$\degree$)  &   2.0             &   2.0             \\
                &   1   &   303.93--305.81  &   0$\farcs$94$\times$0$\farcs$62 (179.0$\degree$)     &   0$\farcs$82$\times$0$\farcs$63 (-8.4$\degree$)  &   2.1             &   2.1             \\
                &   2   &   314.05--315.93  &   0$\farcs$91$\times$0$\farcs$60 (179.2$\degree$)     &   0$\farcs$79$\times$0$\farcs$61 (171.9$\degree$) &   2.5             &   2.4             \\
                &   3   &   315.93--317.81  &   0$\farcs$90$\times$0$\farcs$60 (178.8$\degree$)     &   0$\farcs$79$\times$0$\farcs$61 (171.2$\degree$) &   2.9             &   2.9             \\
        \enddata
    \end{deluxetable*}

    LkCa~15 and MWC~480 were observed in Band 7 during ALMA Cycles 3 $\&$ 4 (project code 2015.1.00657.S). Six correlator setups were designed to provide nearly complete frequency coverage between 275 and 322~GHz. Each spectral setup consisted of 4 Frequency Division Mode (FDM) spectral windows, each with 1920 channels. The channel width was 975~kHz, resulting in a velocity resolution of $\sim$ 1 km s$^{-1}$. Only the first five of these setups had observations taken, resulting in coverage gaps between 306--310~GHz and 318--322~GHz. Observational details including number of antennas, uv-coverage, on-source integration time, and calibrator information are listed in Table \ref{SLS_Table1}. Details of the spectral setups are listed in Table \ref{SLS_Table3}. 
    
    Data was initially calibrated by the ALMA/NAASC staff. Subsequent self-calibration and imaging of the data were completed using \texttt{CASA} 4.3.1. For each execution and disk, the dust continuum was imaged by \texttt{CLEAN}ing the line-free portions of the data with a Briggs robust parameter of -0.5. Three rounds of phase-only self-calibration on the disk continuum emission were then completed with solution intervals of 100s, 30s, and 10s, followed by one iteration of amplitude self-calibration. These calibration tables were applied to the full data for each execution, and the individual spectral windows were then split and continuum subtracted using the CASA task \texttt{uvcontsub} (fit-order~=~1).

\section{Data analysis}
\label{SLS_S3}
\subsection{Initial full-band imaging and spectral extraction}
    \label{SLS_S32}
        Each of the spectral windows for both sources was imaged with \texttt{CLEAN} in CASA 4.3.1, using natural weighting for optimal sensitivity. Resultant synthesized beam sizes and position angles for each spectral window are listed in Table \ref{SLS_Table3}. When a spectral setting had multiple executions taken, the respective measurement sets for each spectral window were concatenated prior to imaging. Each window was initially dirty imaged to determine a characteristic per-channel rms, listed in Table \ref{SLS_Table3}. They were then \texttt{CLEAN}ed down to a threshold of 4$\sigma$, with elliptical \texttt{CLEAN} masks defined for each source to contain the full extent of the gas disk. For MWC~480, the mask had a major axis length of 3$\farcs$5, inclination of 37$\degree$, and PA of 148$\degree$ \citep{Chiang_2001, Pietu_2007, Oberg_2011}. For LkCa~15, the mask had a major axis length of 4$\farcs$5, inclination of 52$\degree$, and PA of 60$\degree$ \citep{Pietu_2007, Isella_2012}. These elliptical masks were applied uniformly across all channels to preserve the unbiased nature of the survey. After imaging, disk-integrated spectra were extracted from the image cubes using the elliptical \texttt{CLEAN} masks. The spectra were averaged where spectral window 3 from setting C and spectral window 0 from setting D overlap.
        
\begin{figure*}[htpb!]
        \centering
        \includegraphics[width=\textwidth]{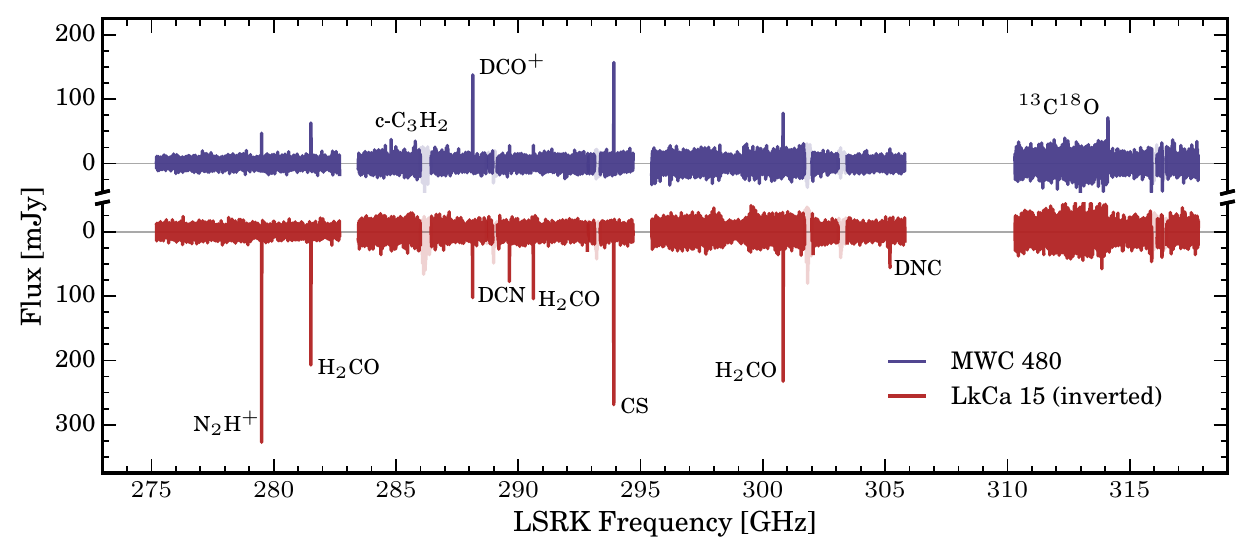}
        \caption{Spectra extracted with the elliptical \texttt{CLEAN} masks for each disk. The spectrum of MWC~480 is shown in blue and that of LkCa~15 is mirrored in red, with spectral regions contaminated by atmospheric absorption lines plotted with transparency. Molecular species with transitions detected at a 4$\sigma$ level are labeled, with the label location denoting within which disk the species is more strongly detected. \label{Figure 3}}
        \end{figure*}

        The full extracted spectra of MWC~480 and LkCa~15 are shown in Figure \ref{Figure 3} in blue and mirrored in red, respectively. Lines visible in the spectra are labeled in the figure, with the location of the label above/below the baseline denoting in which disk the line is stronger. Sections of the spectrum where atmospheric absorption lines are present are plotted with transparency, and the somewhat different sensitivity of each spectral window is visible in the figure.
            
    \subsection{Matched filtering}
    \label{SLS_S31}
\begin{figure*}[htpb!]
        \centering
        \includegraphics[width=\textwidth]{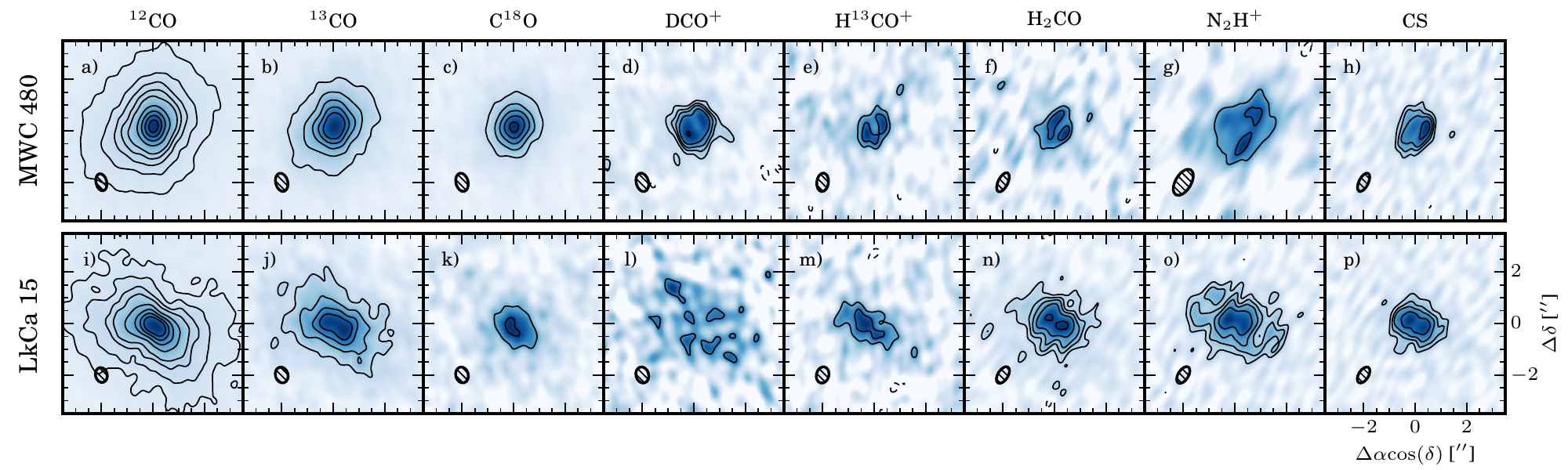}
        \caption{Moment-0 integrated emission maps of observations used as matched filter templates, demonstrating varied emission morphology. Emission for all species has been normalized by dividing by the peak emission. Contours for MWC~480 CO isotopologues (panels a-c) are $[10,20,30,40,50,75,100,125,150]\times\sigma$ and contours for LkCa CO isotopologues (panels i-k) are $[5,10,15,20,30,40,...]\times\sigma$. Contours for all other panels are $[3,5,7,10,15,20,...]\times\sigma$. Synthesized beams are shown in the lower left of each panel. \label{Figure 1}}
        \end{figure*}

        The data were further analyzed by applying a bank of matched filters to attempt to locate any additional weak lines. When the shape of a signal is known (or can be well-approximated), application of a matched filter allows for maximal signal extraction \citep{North_1963}. In \cite{Loomis_2018_MF}, we developed a method to efficiently apply matched filters to interferometric observations, allowing the quick analysis of high bandwidth interferometric spectral surveys and improved SNR. Previous observations of both MWC~480 and LkCa~15 \citep[e.g.,][]{Oberg_2010, Oberg_2011, Huang_2017} have determined the Keplerian rotation pattern of both disks, as well as a variety of radial emission profiles for different molecular species. We use prior observations of $^{12}$CO, $^{13}$CO, C$^{18}$O, DCO$^{+}$, and H$^{13}$CO$^{+}$ \citep{Huang_2017} to generate a bank of data-driven template filters to apply to the line survey data. Filters were also created using strong lines of H$_2$CO, N$_2$H$^+$, and CS imaged from the presently described observations. All of these filters describe identical Keplerian rotation patterns, and thus will likely yield similar results for a given target line, but small differences between the filter responses may provide useful clues about the target line emission morphology, especially if the target line is too weak to image at high resolution. Moment-0 integrated emission maps of each of the filter template molecules are shown in Fig. \ref{Figure 1}, demonstrating the broad range of emission patterns.
        
        Each of the template transitions were imaged with \texttt{CLEAN}, and relatively noise-free approximations of the emission were created by convolving the \texttt{CLEAN} components with their respective restoring beams. These image cubes were then used as data driven filters and applied to each continuum-subtracted spectral window using the \texttt{VISIBLE} package \citep{VISIBLE_ASCL}\footnote{\texttt{VISIBLE} is publicly available under the MIT license at \href{https://github.com/AstroChem/VISIBLE}{https://github.com/AstroChem/VISIBLE} or in the Anaconda Cloud at \href{https://anaconda.org/rloomis/VISIBLE}{https://anaconda.org/rloomis/VISIBLE}}. As the image cubes do not all match the native velocity resolution of our observations, \texttt{VISIBLE} interpolates the filters to match the local velocity resolution of the data being filtered. The interpolated filter is then convolved with the data, producing a normalized filter impulse response spectrum. The technical details of this method can be found in \cite{Loomis_2018_MF}. Spectra from spectral windows that contained multiple executions were first combined via a weighted average, with weights calculated from the rms in each response spectrum, and then normalized after averaging.
        
\begin{figure*}[htpb!]
        \centering
        \includegraphics[width=\textwidth]{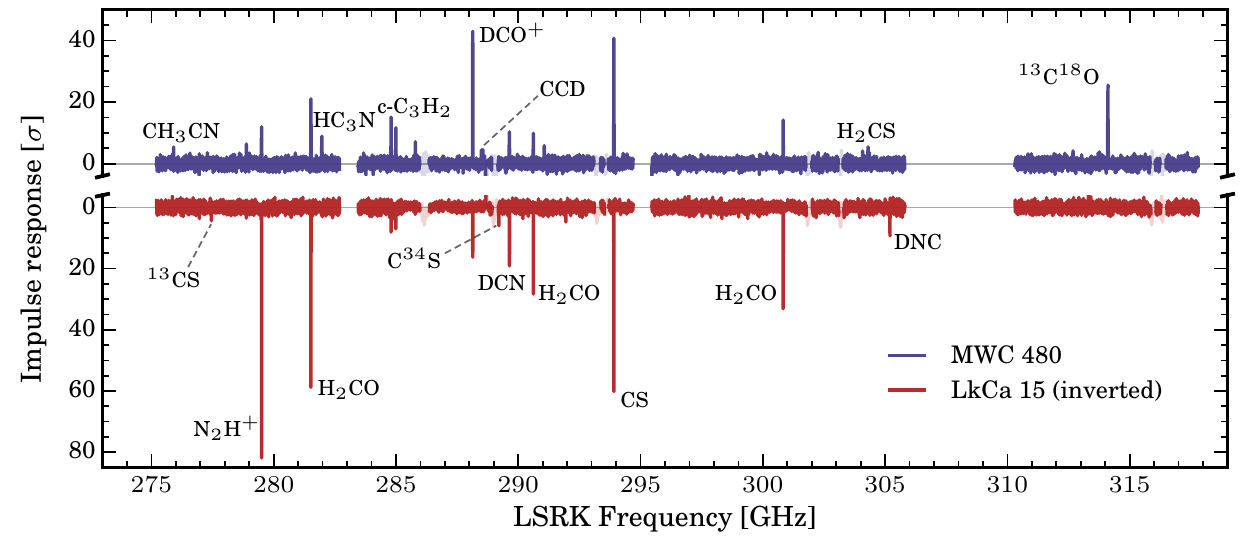}
        \caption{Impulse response spectra for the entire observed bandwidth, produced by filtering the observations with the H$^{13}$CO$^{+}$ template kernel. The spectrum of MWC~480 is shown in blue and that of LkCa~15 is mirrored in red, with spectral regions contaminated by atmospheric absorption lines plotted with transparency. Molecular species with transitions detected at a 4$\sigma$ level are labeled, with the label location denoting in which disk the species is more strongly detected. \label{Figure 2}}
        \end{figure*}

        The full survey filter impulse response spectrum for the H$^{13}$CO$^{+}$ template is shown in Figure \ref{Figure 2} as an illustrative example. Response spectra for the other filter templates produced similar results and are available online as a figure set. The spectrum of MWC~480 is shown in blue and that of LkCa~15 is mirrored in red. As in Figure \ref{Figure 3}, regions contaminated by atmospheric features are plotted with transparency. A comparison with Figure \ref{Figure 3} demonstrates that application of a matched filter results in a higher SNR spectrum. and many weak lines that were previously not seen in Figure \ref{Figure 3} are now visible. Observed lines of the detected species are labeled in the figure, with the location of the label above/below the baseline denoting within which disk the species is more strongly detected. At any given frequency, the sensitivity of both the MWC~480 and LkCa~15 spectra are nearly identical (given the identical integration times, similar uv coverage, similar angular sizes of the disks, and similar linewidths), so a comparison of filter responses between the two disks likely reflects emission strength differences between them. Due to the varied execution time and weather conditions of each setting, however, the individual spectral windows do not have uniform sensitivity and thus we stress that the full normalized spectrum does not portray individual line emission intensity (i.e. the ratio of filter responses of lines in different spectral windows does not reflect the ratio of their total integrated emission). This is especially clear in the relative filter responses of the three strong H$_2$CO lines in LkCa~15: the 4$_{14}$--3$_{13}$ transition at 281.5~GHz and the 4$_{13}$--3$_{12}$ transition at 300.8~GHz have very similar integrated emission, but different filter responses, as the 4$_{13}$--3$_{12}$ transition is in a noisier spectral window. This note aside, the filter response spectra allow for quick identification of all lines detected in the survey.
        
    \subsection{Individual line imaging and flux measurement}
        All spectral lines that were detected by matched filtering in either disk at greater than 4$\sigma$ significance were then individually imaged with \texttt{CLEAN} using natural weighting. Lines that were strongly detected ($>$~20$\sigma$) were imaged at higher spatial resolution. An identical imaging process was used, but with Briggs weighting and a robust value of 0.5. Lower values of robustness produced images with substantially higher noise and thus were not used. The resultant synthesized beam parameters are listed in the respective figure captions in \S\ref{SLS_S4}. Velocity mode in \texttt{CLEAN} was used with a channel width of 1.5 km s$^{-1}$, so that all lines would be on a regular velocity grid. The previously described elliptical \texttt{CLEAN} masks were used for each disk, and the image cubes were \texttt{CLEAN}ed down to a threshold of 4$\sigma$, using the rms values listed in Table \ref{SLS_Table3}. Disk-integrated spectra were then extracted from the image cubes using the \texttt{CLEAN} masks. Fluxes for each transition were determined by integrating these spectra from 2.0 to 9.5 km s$^{-1}$ and 3.5 to 11 km s$^{-1}$ in MWC~480 and LkCa~15, respectively. Uncertainties on each flux measurement were determined through bootstrapping, repeating this process 10000 times on randomly selected nearby emission-free channels, sampled with replacement \citep[see, e.g.][for a description of this technique]{Bergner_2018}. The standard deviations of these values are reported as the uncertainty on the flux measurements.

\section{Results}
\label{SLS_S4}
    \subsection{Overview}
    \label{SLS_S41}
        As shown in Figures \ref{Figure 3} and \ref{Figure 2}, we detect 14 molecular species (including isotopologues) at a $>$4$\sigma$ significance toward MWC~480 and LkCa~15, with 5 of these species (C$^{34}$S, $^{13}$CS, H$_{2}$CS, DNC, and C$_2$D) detected for the first time in a protoplanetary disk (as also presented in \cite{LeGal_2019} for the S-bearing species). 11 species were detected toward MWC~480, with DNC, C$^{34}$S, and $^{13}$CS not detected, and 9 species detected toward LkCa~15, with $^{13}$C$^{18}$O, H$_2$CS, C$_2$D, HC$_3$N, and CH$_3$CN not detected. Observed transitions of the 14 detected species are tabulated in Table \ref{SLS_Table3}. Integrated fluxes (or 2$\sigma$ upper limits for lines not detected at $>$4$\sigma$ with any filter) are listed for each transition, extracted as described in \S\ref{SLS_S32}. The filters that yielded the strongest impulse response for each transition are listed, along with the respective significance of the response (self-filtered responses for the strong lines of H$_2$CO, N$_2$H$^+$, and CS were excluded). From both the above figures and Table \ref{SLS_Table3}, it is clear that the molecular inventories of MWC~480 and LkCa~15 differ dramatically, which might be expected given their different stellar masses, disk masses, radiation environments, and average temperatures. Emission morphologies and their differences are explored on a per molecule basis in the following sub-sections, and then discussed further in \S\ref{SLS_S5}.
        
        
        %

        \begin{deluxetable*}{ccccccccccc}
        \tablecaption{Observed spectral lines\label{SLS_Table4}}
        \tablecolumns{11}
        \tablewidth{\columnwidth}
        \tablehead{             
                                &                                       &                   &           &                   &  \multicolumn{3}{c}{\textbf{MWC~480}} &   \multicolumn{3}{c}{\textbf{LkCa~15}} \\
        \toprule
        	\colhead{Species} & \colhead{Transition} & \colhead{Frequency}  & \colhead{E$_{u}$}  & \colhead{S$_{ij}\mu^{2}$}  & \colhead{Integrated} & \colhead{Filter}  & \colhead{Filter} & \colhead{Integrated}  & \colhead{Filter} & \colhead{Filter}\\[-1ex]
        	&                                       &                   &           &                   & Flux Density      & Response  & & Flux Density  & Response   &     \\
        	&                                       & (MHz)             & (K)       & (D$^2$)           & (mJy km s$^{-1}$) & ($\sigma$) &  & (mJy km s$^{-1}$) & ($\sigma$)  &   }
        	
        	\startdata
            $^{13}$C$^{18}$O    & 3--2                                  & 314199.7$^{a}$    & 30.2      & 0.073$^{b}$       & 422~$\pm$~34      & 25.5              & CS        & 79~$\pm$~31       & $\cdots^{d}$      & $\cdots^{d}$  \\[1ex]
            N$_2$H$^+$          & 3--2                                  & 279511.7          & 26.8      & 335.2           & 228~$\pm$~13      & 17.7              & C$^{18}$O & 1750~$\pm$~20     & 89.8              & C$^{18}$O  \\[1ex]
            H$_{2}$CO           & 4$_{04}$--3$_{03}$                    & 290623.4          & 34.9      & 21.7            & 153~$\pm$~18      & 12.8              & N$_2$H$^+$& 581~$\pm$~25      & 34.4              & C$^{18}$O \\
                                & 4$_{14}$--3$_{13}$                    & 281526.9          & 45.6      & 61.1            & 324~$\pm$~19      & 23.6              & N$_2$H$^+$& 1152~$\pm$~19     & 71.5              & $^{13}$CO \\
                                & 4$_{13}$--3$_{12}$                    & 300836.6          & 47.9      & 47.9            & 342~$\pm$~22      & 16.4              & C$^{18}$O & 1334~$\pm$~34     & 39.4              & C$^{18}$O \\
                                & 4$_{22}$--3$_{21}$                    & 291948.1          & 82.1      & 16.3            & $<$~60$^{c}$      & $\cdots^{d}$      & $\cdots^{d}$    & 80~$\pm$~30       & 4.0           & C$^{18}$O    \\[1ex]
            DCO$^+$             & 4--3                                  & 288143.9$^{a}$    & 34.6      & 189.3$^{b}$     & 784~$\pm$~16      & 46.4              & H$_2$CO   & 502~$\pm$~17      & 26.7              & $^{13}$CO \\[1ex]
            DCN                 & 4--3                                  & 289645.2$^{a}$    & 34.8      & 114.7$^{b}$     & 104~$\pm$~19      & 10.3              & CS        & 410~$\pm$~22      & 23.6              & N$_2$H$^+$ \\[1ex]
            DNC                 & 4--3                                  & 305206.2$^{a}$    & 36.6      & 37.2$^{b}$      & $<$~32$^{c}$      & 3.6               & H$^{13}$CO$^+$& 228~$\pm$~22  & 15.9              & $^{13}$CO  \\[1ex]
            CS                  & 6--5                                  & 293912.1          & 49.4      & 22.9            & 811~$\pm$~19      & 41.0              & DCO$^+$   & 1497~$\pm$~19     & 62.9              & H$_2$CO  \\[1ex]
            C$^{34}$S           & 6--5                                  & 289209.1          & 38.2      & 22.2            & $<$~32$^{c}$      & $\cdots^{d}$      & $\cdots^{d}$& 90~$\pm$~19     & 6.2               & H$_2$CO \\[1ex]
            $^{13}$CS           & 6--5                                  & 277455.4          & 46.6      & 23.0            & $<$~28$^{c}$      & $\cdots^{d}$      & $\cdots^{d}$& 61~$\pm$~17     & 4.4               & H$_2$CO \\[1ex]
            H$_{2}$CS           & 8$_{17}$--7$_{16}$                    & 278886.4          & 73.4      & 64.1            & 83~$\pm$~19       & 6.4               & CS        & $<$~40$^{c}$      & 3.5               & CS  \\
                                & 9$_{19}$--8$_{18}$                    & 304306.0          & 86.2      & 72.3            & 51~$\pm$~18       & 5.4               & CS        & $<$~38$^{c}$      & $\cdots^{d}$      & $\cdots^{d}$ \\
                                & 9$_{18}$--8$_{17}$                    & 313714.9          & 88.5      & 72.3            & 80~$\pm$~41       & 4.2               & H$^{13}$CO$^+$& 52~$\pm$~45      & 4.2      & H$^{13}$CO$^+$ 
                                \\[1ex]
            c-C$_3$H$_2$        & 8$_{18}$--7$_{07}$                    & 284805.2$^{e}$    & 64.3      & 239.2           & 166~$\pm$~23      & 15.1              & CS        & 104~$\pm$~32      & 8.1               & CS \\
                                & 7$_{16}$--6$_{25}$                    & 284998.0$^{f}$    & 61.2      & 174.0           & 143~$\pm$~21      & 11.7              & CS        & 101~$\pm$~28      & 7.0               & CS \\
                                & 6$_{34}$--5$_{23}$                    & 285795.7          & 54.7      & 110.8           & 118~$\pm$~24      & 7.3               & H$^{13}$CO$^+$ & $<$~68$^{c}$ & $\cdots^{d}$      & $\cdots^{d}$  \\[1ex]
            C$_2$D              & 4--3 J=$\frac{9}{2}$--$\frac{7}{2}$   & 288499.0$^{a}$    & 34.6      & 7.6$^{b}$       & 63~$\pm$~22       & 4.5               & DCO$^+$   & $<$~44$^{c}$      & $\cdots^{d}$      & $\cdots^{d}$   \\
                                & 4--3 J=$\frac{7}{2}$--$\frac{5}{2}$   & 288554.6$^{a}$    & 34.6      & 5.7$^{b}$       & 84~$\pm$~16       & 4.9               & DCO$^+$   & $<$~40$^{c}$      & $\cdots^{d}$      & $\cdots^{d}$ \\[1ex]
            HC$_3$N             & 31-30                                 & 281976.8$^{a}$    & 216.5     & 431.7$^{b}$     & 107~$\pm$~16      & 8.9               & CS        & 44~$\pm$~18       & $\cdots^{d}$      & $\cdots^{d}$  \\
                                & 32-31                                 & 291068.4$^{a}$    & 230.5     & 445.6$^{b}$     & 85~$\pm$~19       & 5.8               & CS        & 64~$\pm$~21       & $\cdots^{d}$      & $\cdots^{d}$  \\
                                & 33-32                                 & 300159.7$^{a}$    & 244.9     & 459.6$^{b}$     & 102~$\pm$~22      & 3.6               & DCO$^+$   & $<$~56$^{c}$      & $\cdots^{d}$      & $\cdots^{d}$  \\[1ex]
            CH$_3$CN            & 15$_{0}$--14$_{0}$                    & 275915.6$^{a}$    & 105.9     & 1700$^{b}$      & 60~$\pm$~20       & 5.4               & CS        & $<$~32$^{c}$      & $\cdots^{d}$      & $\cdots^{d}$  \\
                                & 16$_{0}$--15$_{0}$                    & 294302.4$^{a}$    & 120.1     & 1814$^{b}$      & 50~$\pm$~10       & 3.2               & CS        & $<$~38$^{c}$      & $\cdots^{d}$      & $\cdots^{d}$  \\
            \enddata
        
        \tablenotetext{}{$^{a}$ Center frequency of collapsed hyperfine components (spacing smaller than channel width).}
        \tablenotetext{}{$^{b}$ S$_{ij}\mu^{2}$ of combined hyperfine components.}
        \tablenotetext{}{$^{c}$ Upper limits are 2$\sigma$.}
        \tablenotetext{}{$^{d}$ $\cdots$ denotes transition not detected above 3$\sigma$ with any filter}
        \tablenotetext{}{$^{e}$ c-C$_3$H$_2$ 8$_{18}$--7$_{07}$ is spectrally coincident with c-C$_3$H$_2$ 8$_{08}$--7$_{17}$}
        \tablenotetext{}{$^{f}$ c-C$_3$H$_2$ 7$_{16}$--6$_{25}$ is partially blended with c-C$_3$H$_2$ 7$_{26}$--6$_{15}$}
        \end{deluxetable*}


    \subsection{CO isotopologues: $^{13}$C$^{18}$O}
    \label{SLS_S42}
        \begin{figure*}[ht!]
        \centering
        \includegraphics[width=\textwidth]{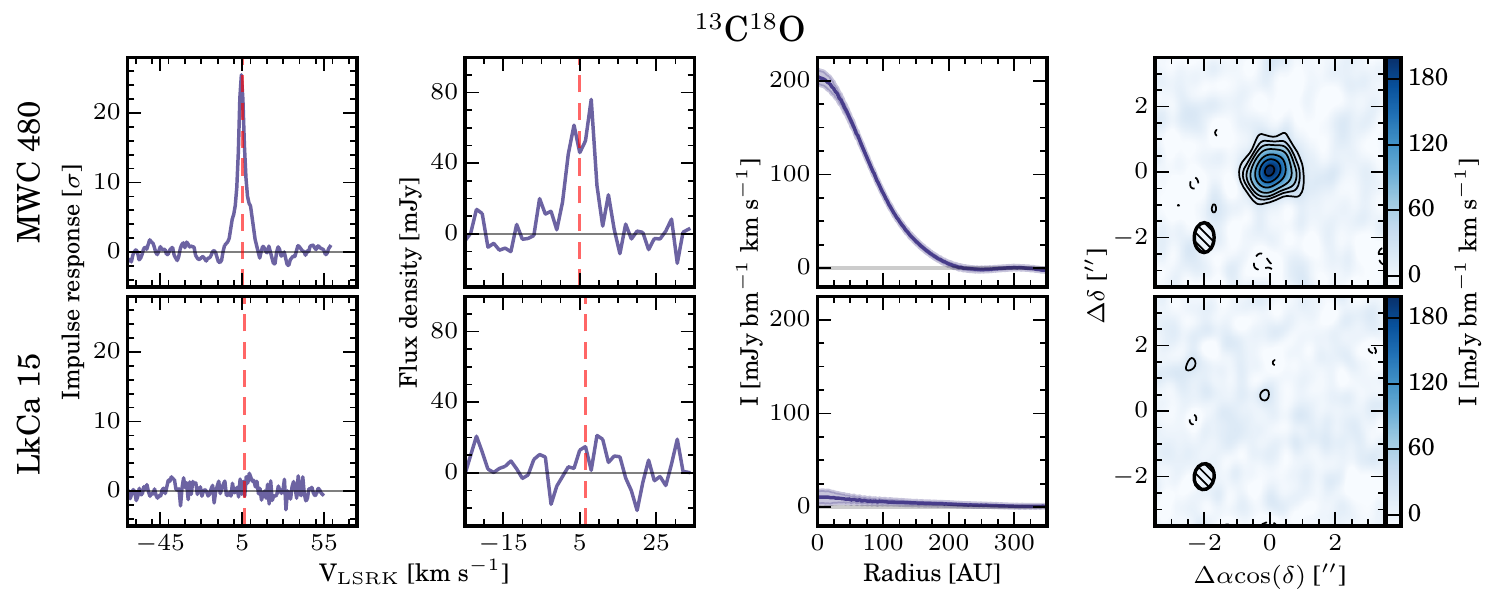}
        \caption{$^{13}$C$^{18}$O observations in MWC~480 (top) and LkCa~15 (bottom). \textit{Far left:} Matched filter impulse response spectra using the filter which yielded the strongest response (see Table \ref{SLS_Table3}). Source systemic velocity is denoted by the dashed red line. \textit{Middle left}: Aperture extracted spectra. Source systemic velocity is denoted by the dashed red line. \textit{Middle right}: Deprojected and azimuthally-averaged radial profiles. 1$\sigma$ uncertanties are shown in shaded blue. \textit{Far right}: Moment-0 maps showing total integrated $^{13}$C$^{18}$O emission. All contours are $[-3,3,5,7,10,15,20,25,...]\times\sigma$. 
        \label{13C18O Fig}}
        \end{figure*}
        
        Figure \ref{13C18O Fig} shows that $^{13}$C$^{18}$O 3--2 emission is strongly detected ($>$20~$\sigma$) in MWC~480 but not observed in LkCa~15. The emission in MWC~480 is compact and centrally peaked, similar to the C$^{18}$O emission shown in Figure \ref{Figure 1}. If $^{13}$C$^{18}$O is optically thin, its radial extent may directly trace the CO snowline \citep[e.g.][]{Zhang_2017}. By deprojecting and azimuthally averaging the emission and then deconvolving the beam, we find that the emission extends out to $\sim$120~AU. This is roughly in agreement with predictions from the temperature profile used in \cite{Oberg_2015}, where a 20~K snowline would be at $\sim$135~AU. A more detailed investigation of the midplane temperature profile of MWC~480 is beyond the scope of this paper, but the high SNR of the $^{13}$C$^{18}$O detection suggest that an analysis of the CO isotopologues similar to that presented in \cite{Zhang_2017} is feasible for MWC~480.

    \subsection{Species sensitive to CO freeze-out: N$_2$H$^+$ $\&$ H$_2$CO}
    \label{SLS_S43}
        \begin{figure*}[ht!]
        \centering
        \includegraphics[width=\textwidth]{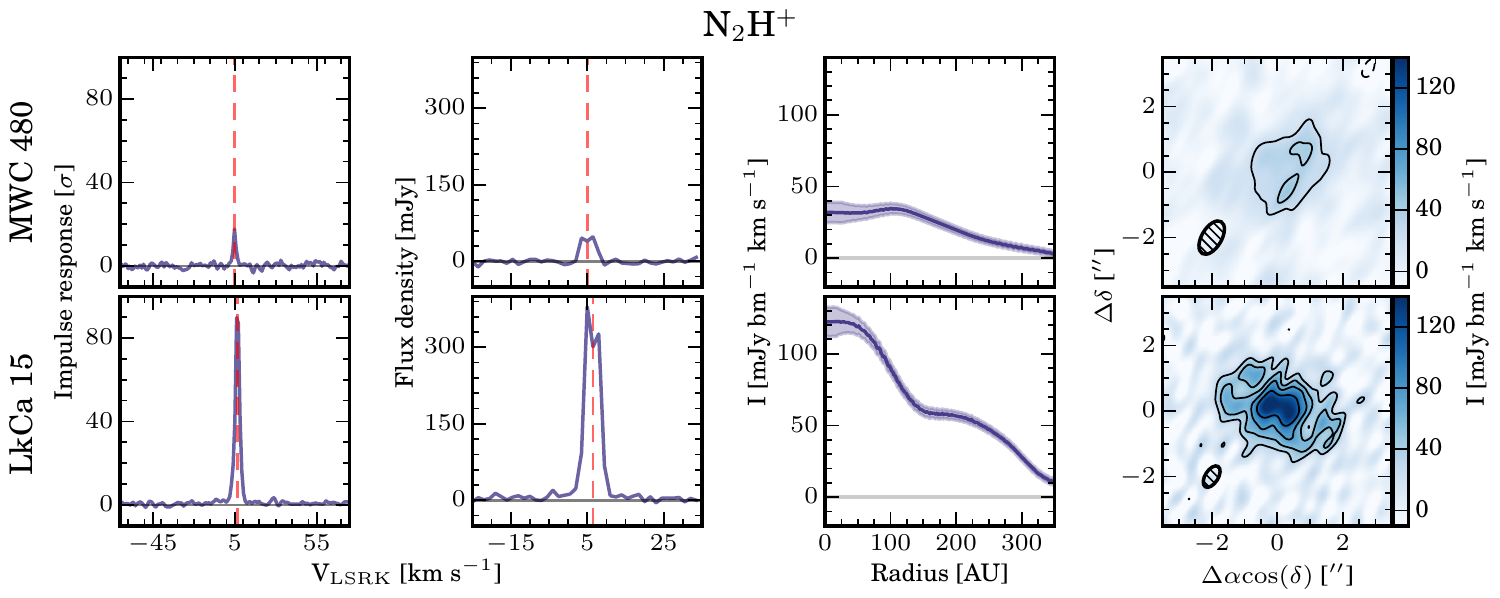}
        \includegraphics[width=\textwidth]{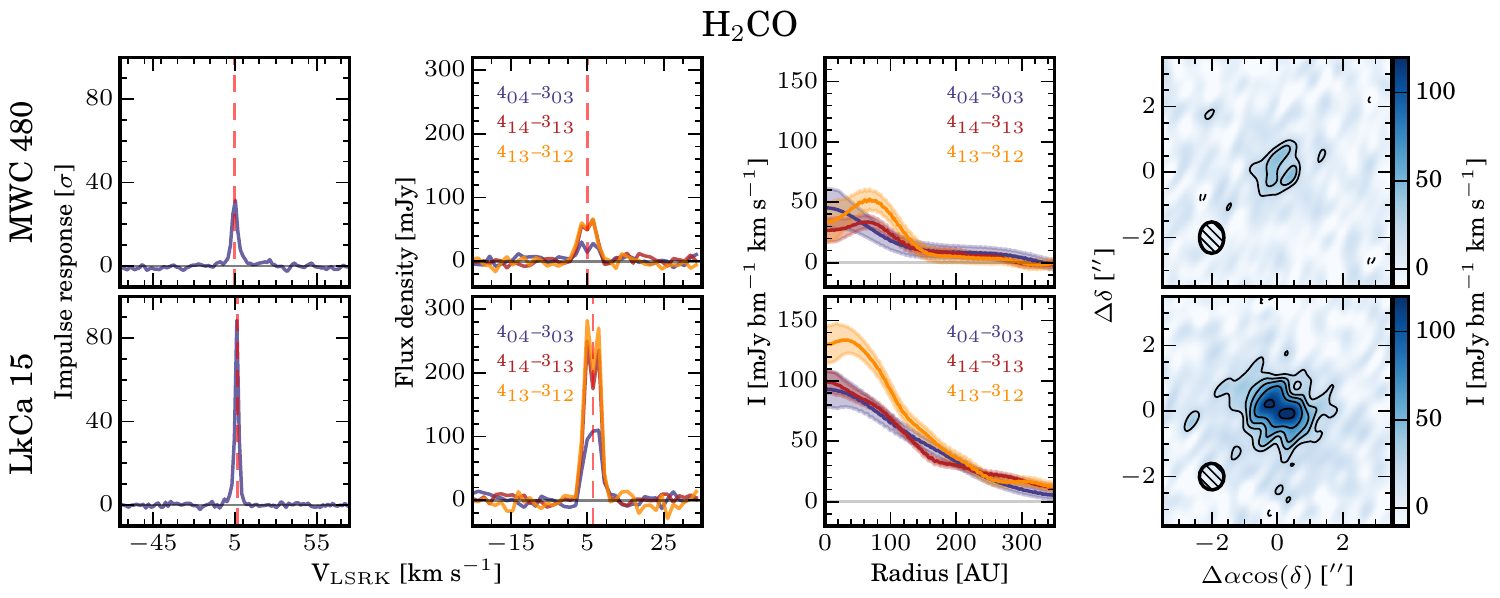}
        \caption{N$_2$H$^+$ (top) and H$_{2}$CO (bottom) observations in MWC~480 and LkCa~15. Panel descriptions are identical to Figure \ref{13C18O Fig}. The H$_2$CO impulse response and moment-0 panels show stacked data, while the aperture extracted spectra and radial profiles are shown for each individual transition. The same convention is used for all subsequent figures. \label{CO_Snow_Fig}}
        \end{figure*}
    
        N$_2$H$^+$ 3--2 emission is strongly detected toward both disks (Figure \ref{CO_Snow_Fig}), but is much brighter toward LkCa~15 (flux ratio of $\sim$7.7; 1750~$\pm$~20 vs 228~$\pm$~13 mJy km s$^{-1}$, respectively). Without detailed chemical modeling, it is unclear whether this is due to effects such as variations in total elemental abundances or ionization between the disks, or enhanced destruction in MWC~480 due to a higher disk temperature and less CO freeze-out. Similar to $^{13}$C$^{18}$O, N$_2$H$^+$ has also been proposed as a tracer of the CO snowline \citep[e.g.,][]{Qi_2013_H2CO, Qi_2013_Sci, Qi_2015}, although its simplicity as a tracer of the snowline location has been debated \citep{vant_Hoff_2017}. Reaction with CO is one of the main destruction pathways of N$_2$H$^+$, and the inner edge of its emission ring may therefore trace the midplane CO abundance profile. The N$_2$H$^+$ emission in both disks demonstrate this ringed morphology, although the emission around LkCa~15 displays a double ring. A similar double ring morphology has been observed for both DCN and H$^{13}$CO$^{+}$ in LkCa~15 \citep{Huang_2017}. By deprojecting and azimuthally averaging the emission from both disks, we find an inner ring radius of $\sim$115~AU for MWC~480 and $\sim$40~AU for LkCa~15. The second ring of emission in LkCa~15 peaks at $\sim$220~AU. The N$_2$H$^+$ inner ring radius for MWC~480 is similar to the radial edge of the $^{13}$C$^{18}$O emission, supporting the hypothesis that they are both tracing the CO snowline location.

        Multiple lines of H$_{2}$CO are strongly detected toward both disks (Figure \ref{CO_Snow_Fig}), but as with N$_2$H$^{+}$ the emission is much brighter toward LkCa~15 (flux ratios of 3.5-3.9). Grain surface formation of H$_{2}$CO through sequential hydrogenation of CO is thought to be a major contributor to H$_{2}$CO abundances in disks \citep{Watanabe_2002, Fuchs_2009, Qi_2013_H2CO, Loomis_2015, Carney_2017, Oberg_2017}. Similar logic may hold as with N$_2$H$^+$ then, where CO freezeout in the colder LkCa~15 disk enhances H$_{2}$CO formation compared with the warmer MWC~480 disk. The H$_{2}$CO in LkCa~15 also has an emission profile that is suggestive of a possible double ring, similar to that seen for N$_2$H$^+$. The detection of many lines of H$_2$CO spanning a wide range of upper state energies (35--82 K) enables the use of rotational diagrams to constrain the H$_2$CO excitation temperature \citep[e.g.][]{Carney_2017}, and further analysis of these observations are presented within the context of a larger survey of H$_{2}$CO in protoplanetary disks in Pegues et al. (in pre).

    \subsection{Deuterated species: DCO$^+$, DCN, $\&$ DNC}
    \label{SLS_S44}
        \begin{figure*}[ht!]
        \centering
        \includegraphics[width=\textwidth]{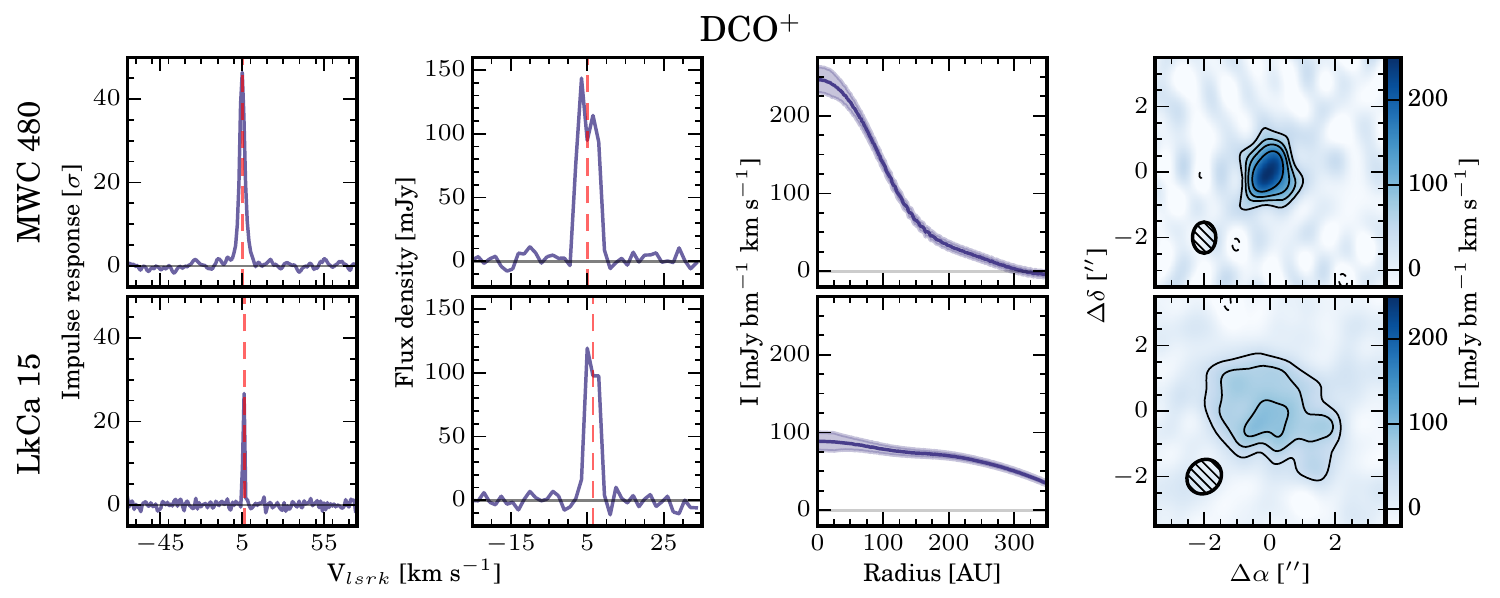}
        \includegraphics[width=\textwidth]{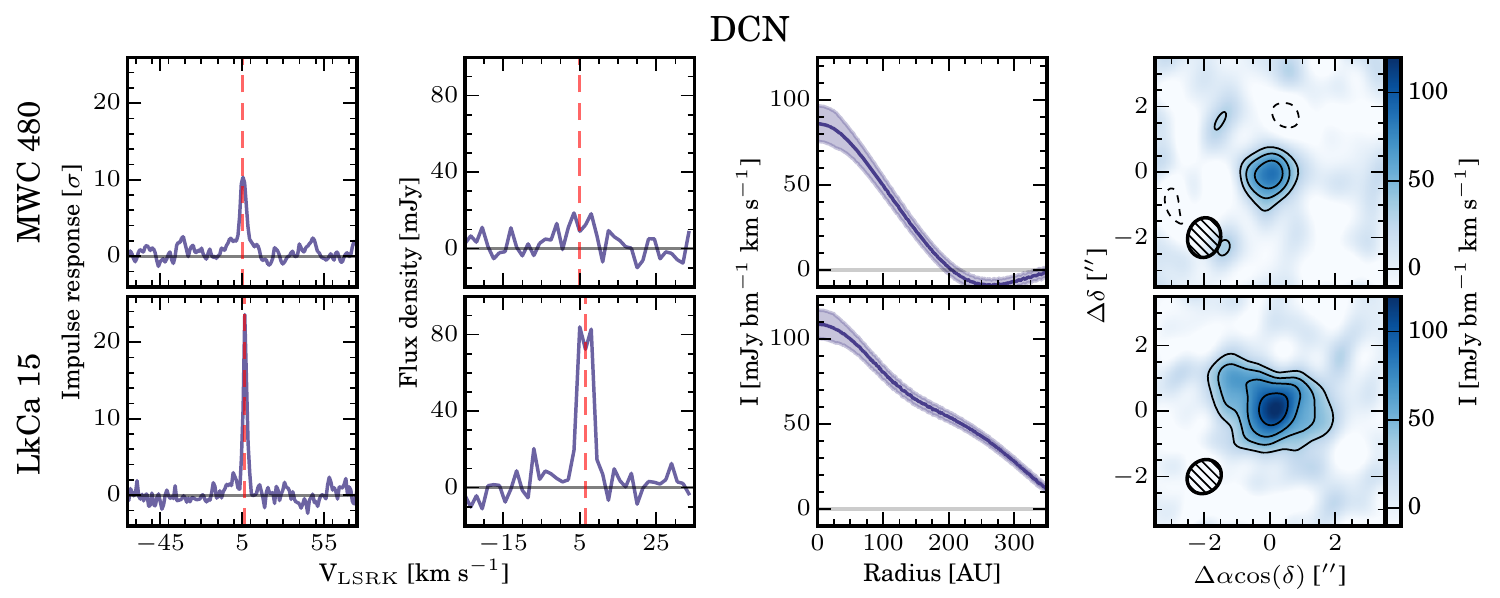}
        \includegraphics[width=\textwidth]{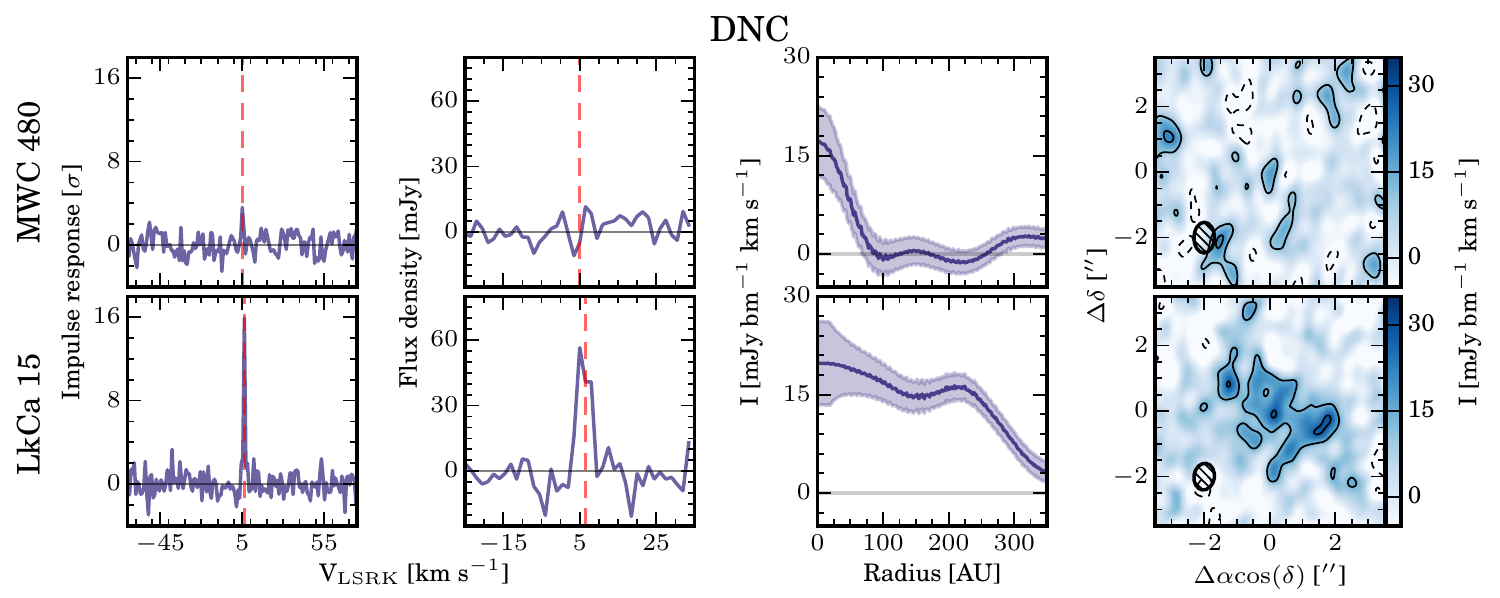}
        \caption{DCO$^+$, DCN, and DNC observations in MWC~480 and LkCa~15. Panel descriptions are identical to Figure \ref{13C18O Fig}, but contours for DNC are $[-2,2,4,6,8,...]\times\sigma$. \label{Deut Fig}}
        \end{figure*}

        An inventory of the deuterium chemistry of a small sample of protoplanetary disks, including both MWC~480 and LkCa~15, was initially explored in \cite{Huang_2017}, tracing the J=3--2 transitions of both DCO$^+$ and DCN, as well as their $^{13}$C isotopologues. Here we observe the J=4--3 transitions of these species and find very similar results. As in \cite{Huang_2017}, we observe bright compact emission of DCO$^+$ in MWC~480 and more diffuse emission in LkCa~15 (Figure \ref{Deut Fig}). \cite{Huang_2017} report an inner hole in the MWC~480 DCO$^{+}$ emission, but we do not have the spatial resolution to confirm such a feature. The DCO$^{+}$ radial profile around LkCa~15 is suggestive of a double ring morphology similar to that seen in N$_2$H+.
        
        DCN is detected toward both disks with emission toward LkCa~15 being 4.1$\times$ stronger (Figure \ref{Deut Fig}), consistent with the flux ratio of $\sim$4 reported in \cite{Huang_2017}. Similar emission profiles are also observed, with emission in MWC~480 being more centrally peaked and compact, and emission in LkCa~15 containing a central peak and an outer ring.


        
        We additionally report the first detection of DNC in a protoplanetary disk, with strong emission detected around LkCa~15 (Figure \ref{Deut Fig}). The emission appears to have a double ring morphology similar to DCN and DCO$^{+}$. In contrast to these two species, however, the ratio of the outer ring surface brightness to the inner ring surface brightness is considerably higher for DNC. A similar difference in the emission morphologies of HCN and HNC has been observed in the TW Hya and HD 163296 disks \citep{Graninger_2015}.

        

    \subsection{Sulfur bearing species: CS, C$^{34}$S, $^{13}$CS, $\&$ H$_2$CS}
    \label{SLS_S45}
        \begin{figure*}[p!]
        \centering
        \includegraphics[width=0.75\textwidth]{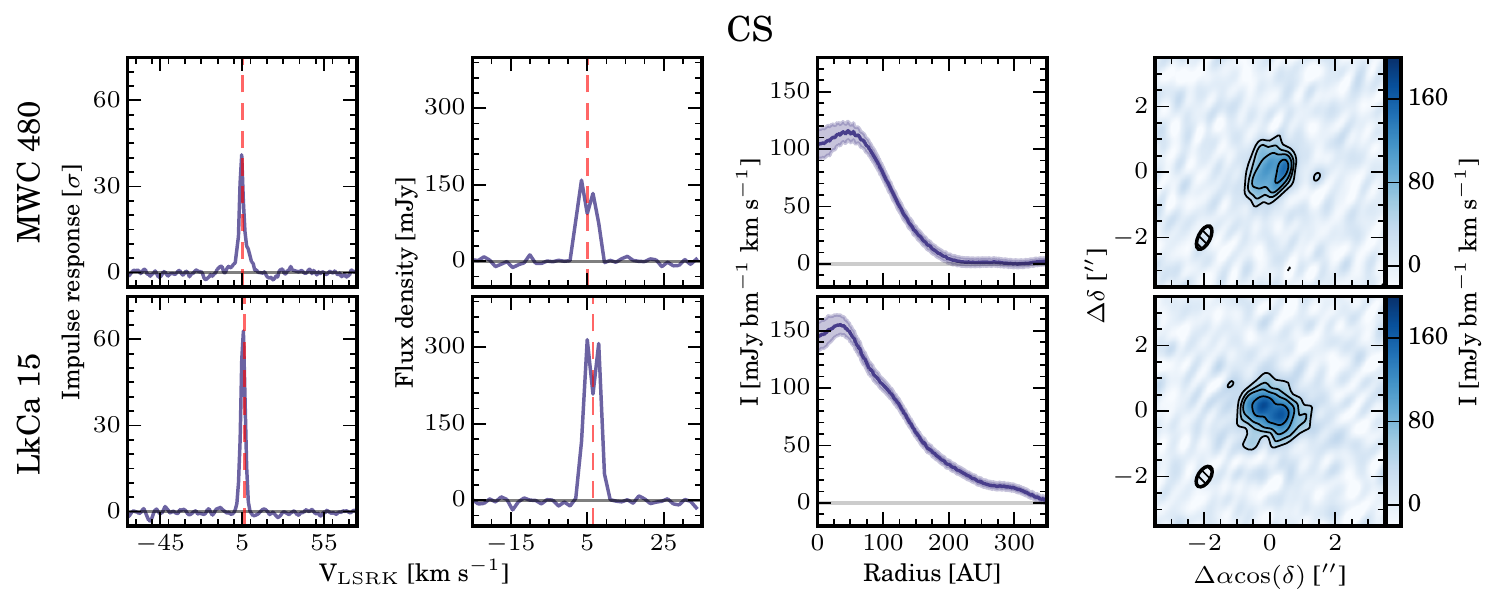}
        \includegraphics[width=0.75\textwidth]{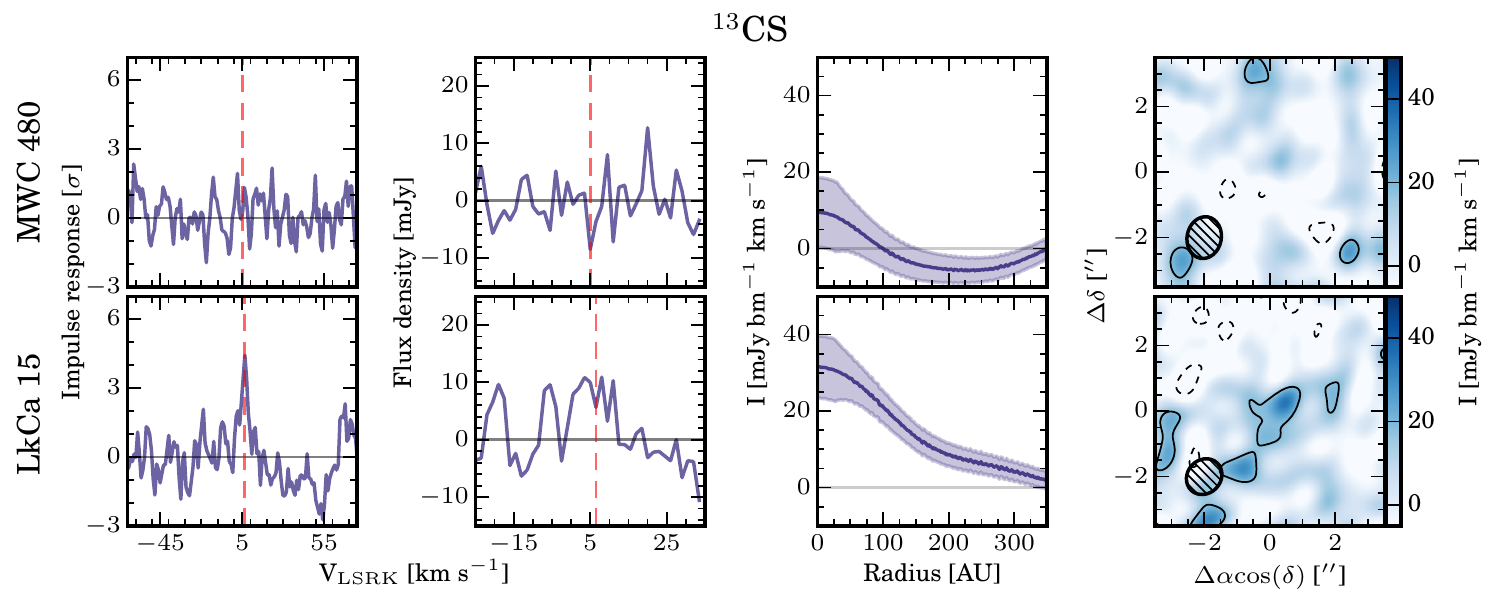}
        \includegraphics[width=0.75\textwidth]{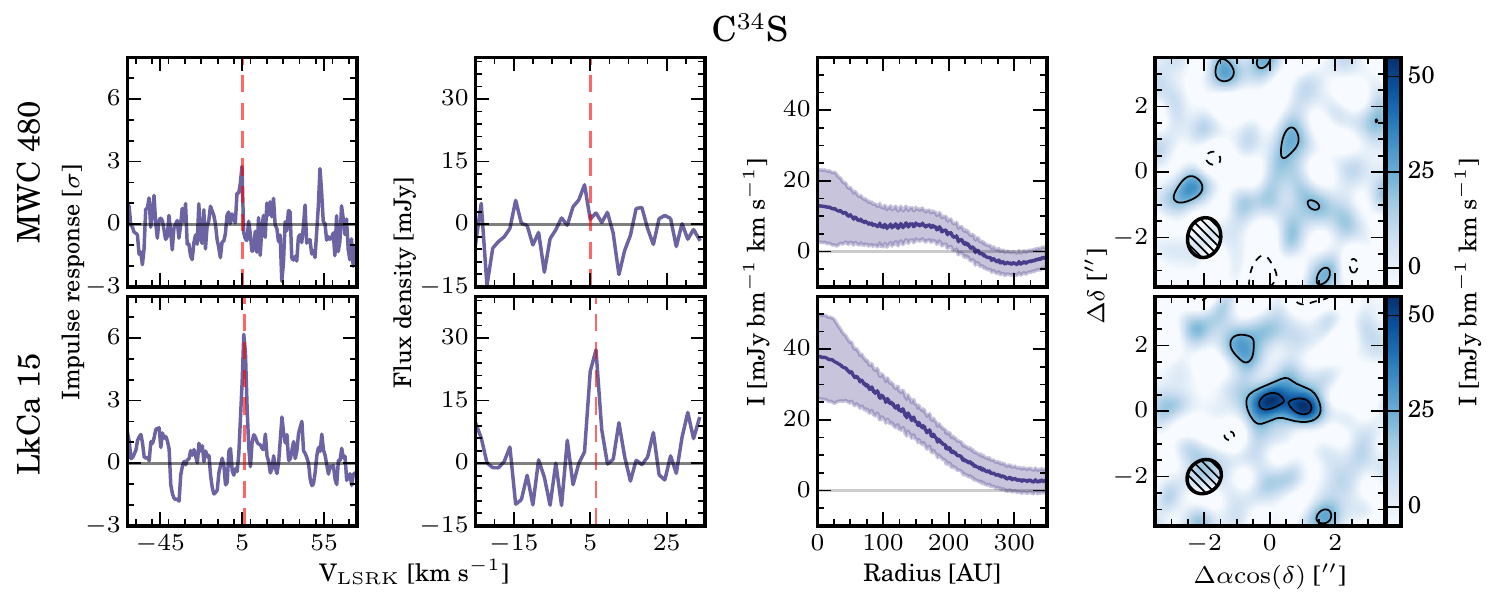}
        \includegraphics[width=0.75\textwidth]{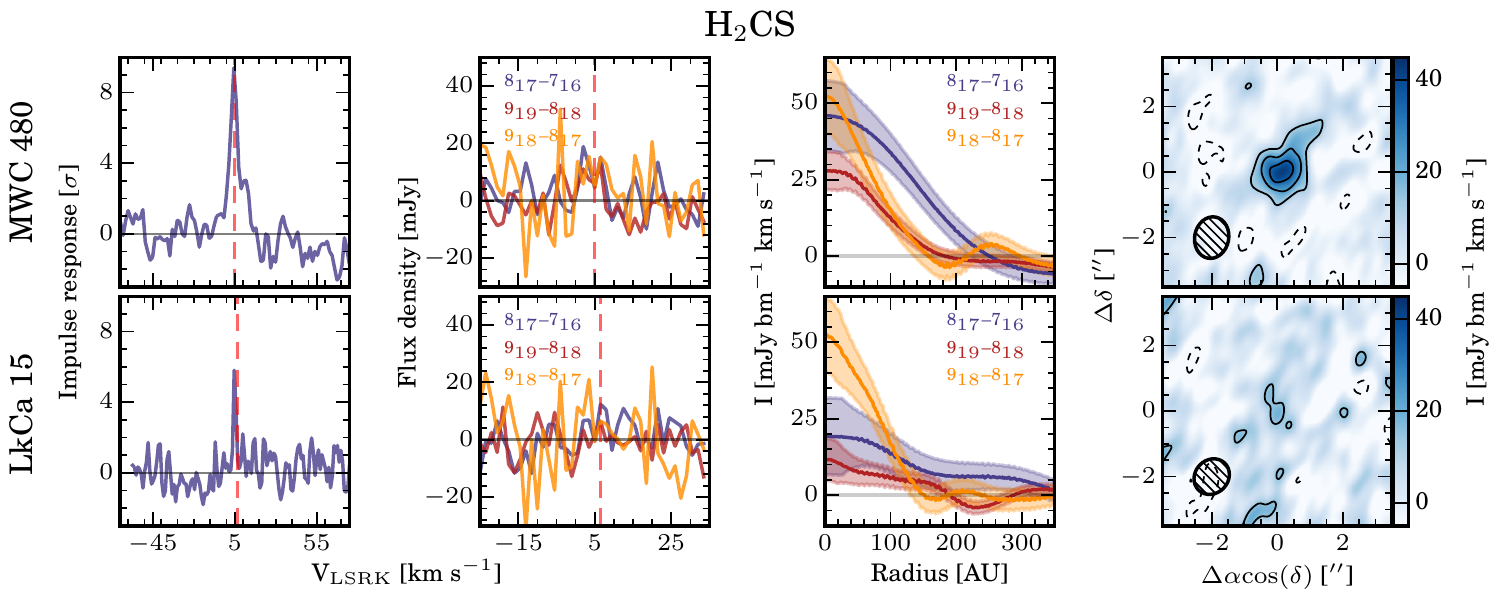}
        \caption{CS, C$^{34}$S, $^{13}$CS, and H$_2$CS observations in MWC~480 and LkCa~15. Panel descriptions are identical to Figure \ref{13C18O Fig}, but C$^{34}$S, $^{13}$CS, and H$_2$CS have contours of $[-2,2,4,6,8,...]\times\sigma$. 
        \label{Sulfur Fig}}
        \end{figure*}
        
        CS 6--5 emission is strongly detected toward both disks (Figure \ref{Sulfur Fig}), but is brighter toward LkCa~15 (flux ratio of $\sim$1.8; 1497~$\pm$~19 vs 811~$\pm$~19 mJy km s$^{-1}$, respectively), with both the $^{13}$CS and C$^{34}$S isotopologues additionally detected toward LkCa~15. The emission morphology of all CS isotopologues in both disks is compact, but the main isotopologue shows a small dip in the emission center toward MWC~480, similar to the CS emission distribution toward TW Hya \citep{Teague_2017}. The central dip in CS emission toward LkCa~15 can likely be explained by its large gas and dust cavity \citep[$\sim$45 AU in CO][]{Jin_2019}. 
        
        In addition to the CS isotopologues, three lines of H$_2$CS (8$_{17}$--7$_{16}$, $_{19}$--8$_{18}$, and 9$_{18}$--8$_{17}$) are detected toward MWC~480 with a very compact emission distribution. No emission is detected toward LkCa~15 above a 4$\sigma$ level. The SNR of the image plane detection in MWC~480 is quite low, but the stacked filter response is $>$8$\sigma$. We briefly discuss the formation chemistry of H$_2$CS in \S\ref{SLS_S5} as well as in \cite{LeGal_2019}, along with possible explanations for why it is detected in MWC~480 and not in LkCa~15.


    \subsection{Hydrocarbons: c-C$_3$H$_2$ $\&$ C$_2$D}
    \label{SLS_S46}
\begin{figure*}[p!]
        \centering
        \includegraphics[width=\textwidth]{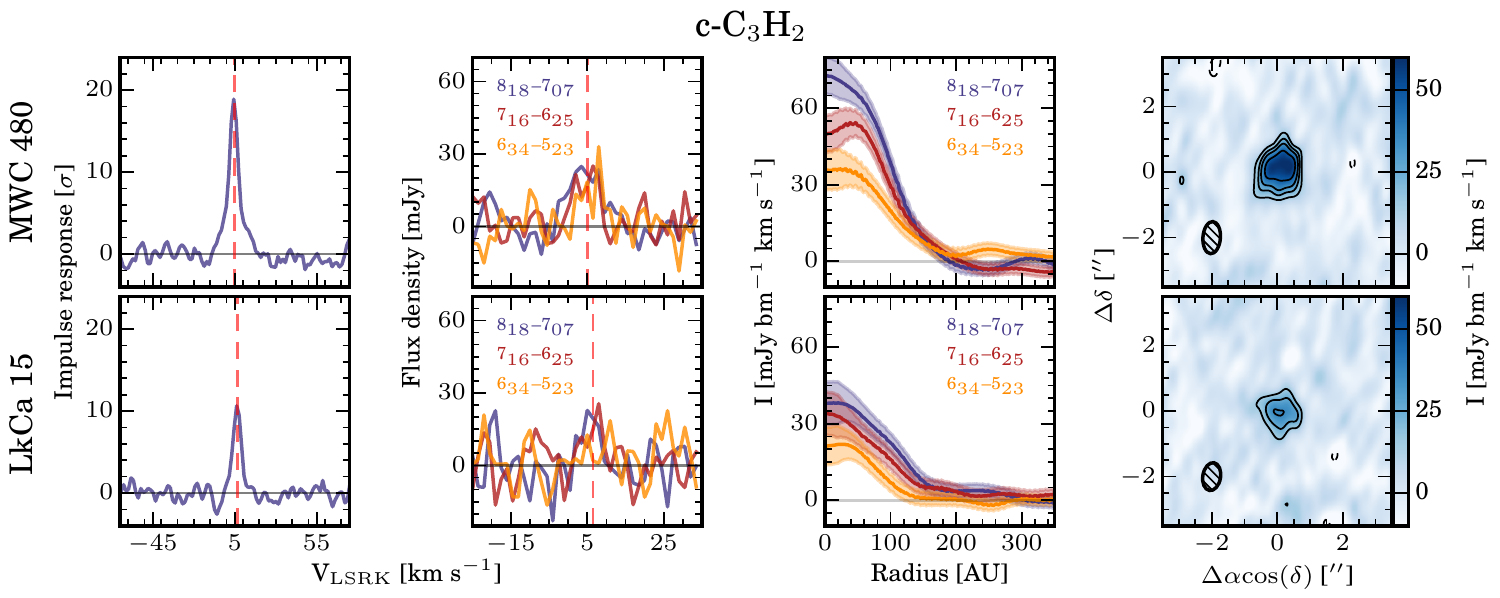}
        \includegraphics[width=\textwidth]{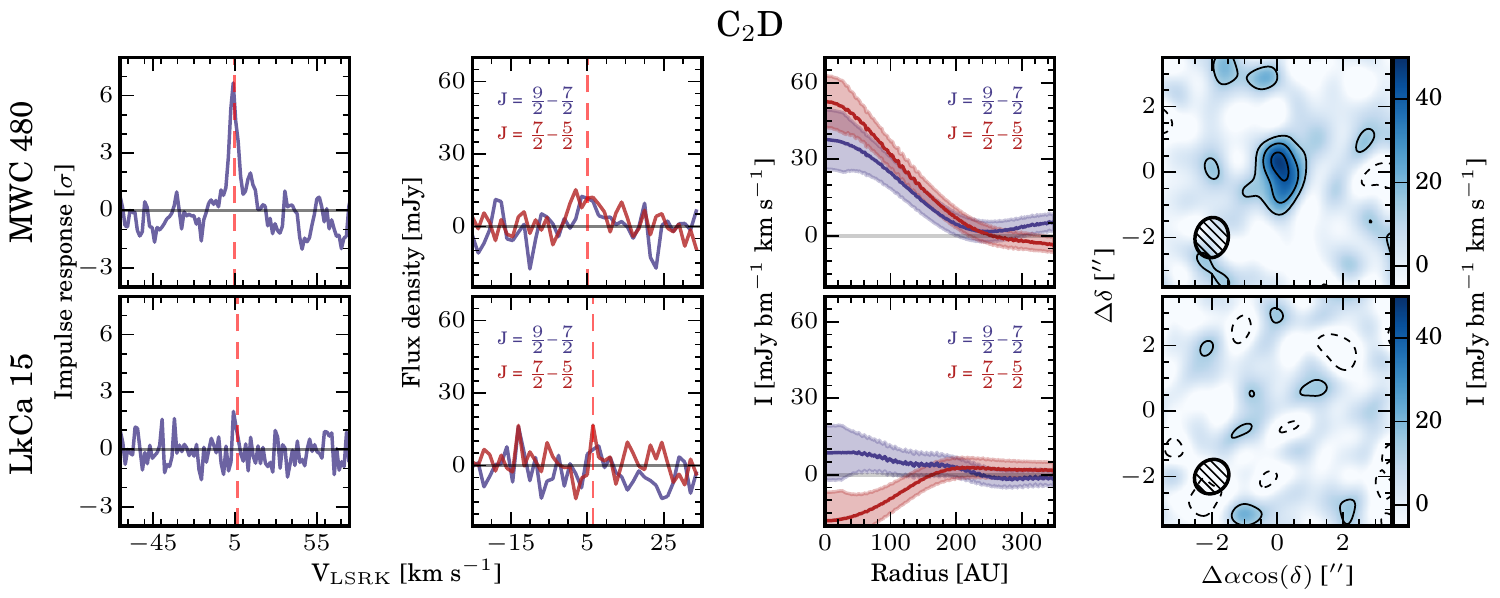}
        \caption{c-C$_3$H$_2$ and C$_2$D observations in MWC~480 and LkCa~15. Panel descriptions are identical to Figure \ref{13C18O Fig}, but contours for C$_2$D are $[-2,2,4,6,8,...]\times\sigma$.
        \label{Hydrocarbon Fig}}
        \end{figure*}

        Numerous transitions of the hydrocarbon ring c-C$_3$H$_2$ are detected around both disks, with emission being much stronger toward MWC~480 (Figure \ref{Hydrocarbon Fig}). This species has previously been reported in abundance toward both HD 163296 \citep{Qi_2013_C3H2} and TW Hya \citep{Bergin_2016}, where the emission had clear ringed distributions. Our spatial resolution is too low to resolve the c-C$_3$H$_2$ radial distribution, but the MWC~480 radial profile hints at a ring.
    
        Emission rings have also been observed for the hydrocarbon C$_2$H \citep{Kastner_2015, Bergin_2016}, although more varied emission morphologies have recently been observed in other disks \citep{Cleeves_2018, Bergner_2019}. C$_2$H was included in our original survey plan, but unfortunately all transitions were within the unobserved spectral setting. We do, however, detect its deuterated isotopologue C$_2$D towards MWC~480 (Figure \ref{Hydrocarbon Fig}). Once again, our spatial resolution is too low to determine a detailed emission profile, but the detection of this deuterated hydrocarbon presents the opportunity for future comparative analysis with other deuterated molecular tracers such as DCN and DCO$^{+}$. The filter with the highest response to the C$_2$D line is DCO$^{+}$, tentatively suggesting that they have similar morphologies.

    \subsection{Nitriles: HC$_3$N $\&$ CH$_3$CN}
    \label{SLS_S47}
        Similar to the hydrocarbon species, the nitriles HC$_3$N and CH$_3$CN are detected in MWC~480 but not in LkCa~15, although there may be a marginal detection of HC$_3$N toward LkCa~15. The emission is spatially compact for both species. These data are presented and discussed in \cite{Bergner_2018} in the context of a larger survey of HC$_3$N and CH$_3$CN in protoplanetary disks.
        
        \begin{figure*}[ht!]
        \centering
        \includegraphics[width=\textwidth]{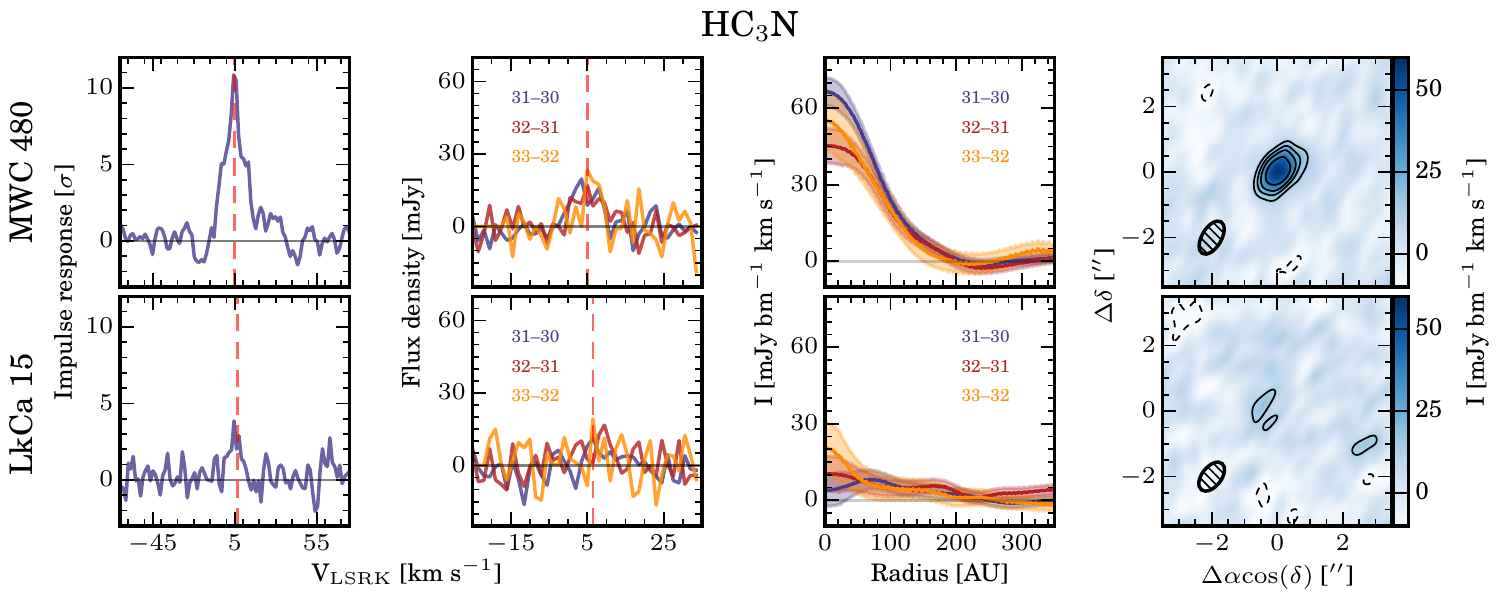}
        \includegraphics[width=\textwidth]{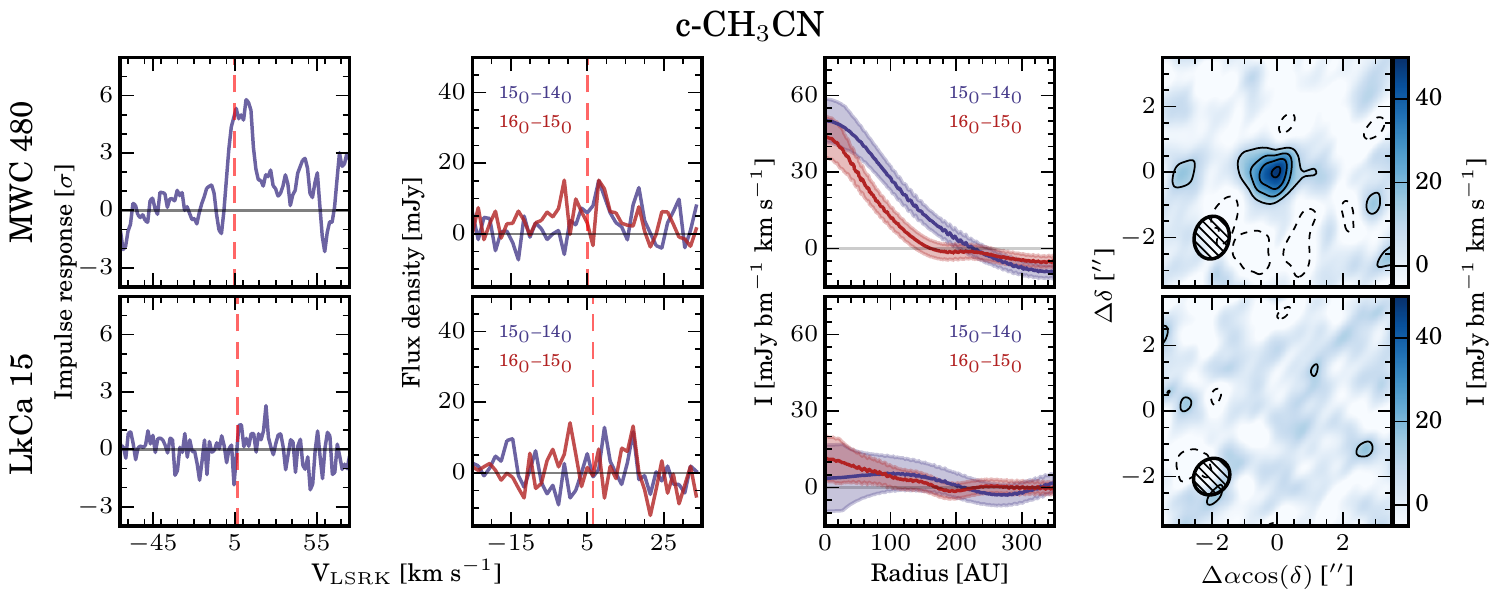}
        \caption{HC$_3$N and CH$_{3}$CN observations in MWC~480 and LkCa~15. Panel descriptions are identical to Figure \ref{13C18O Fig}, but with contours of $[-2,2,4,6,8,...]\times\sigma$.
        \label{Nitriles Fig}}
        \end{figure*}

\section{Discussion}
\label{SLS_S5}
    \subsection{Comparative analysis}
    \label{SLS_S51}
\begin{figure*}[ht!]
        \centering
        \includegraphics[width=\textwidth]{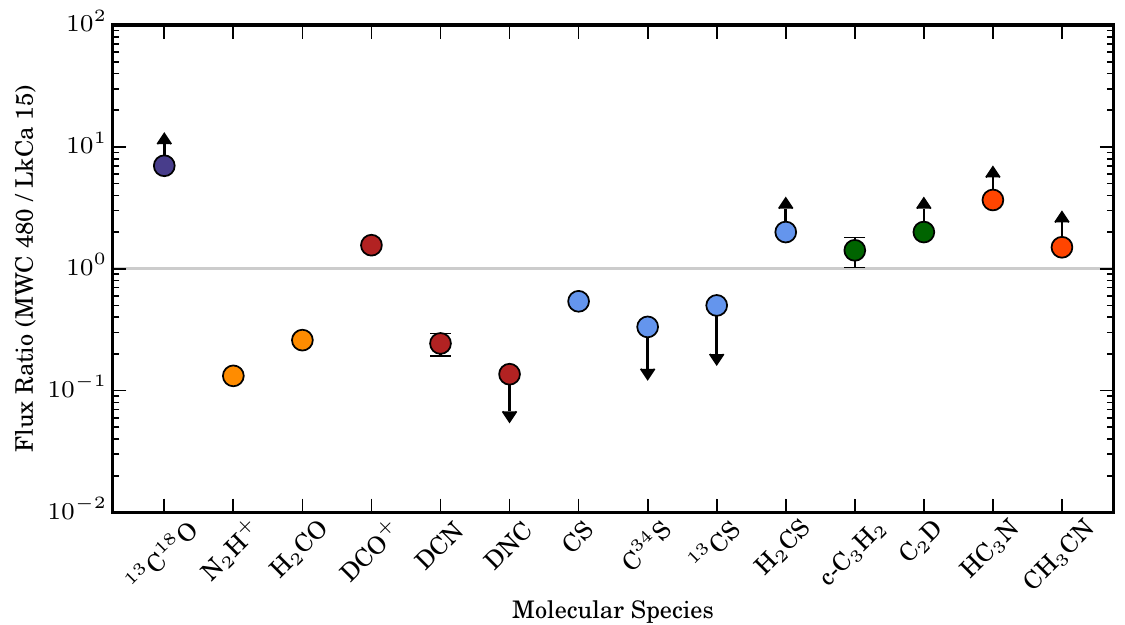}
        \caption{Integrated flux density ratios between MWC~480 and LkCa~15 for each molecule. When a molecule is not detected in one disk, ratio limits are calculated using the 2$\sigma$ upper limits presented in Table 3. Molecules are color coded by the molecular groups used in \S\ref{SLS_S4}.
        \label{Ratio Fig}}
        \end{figure*}

        To compare the molecular inventories of MWC~480 and LkCa~15, we plot integrated flux density ratios in Fig. \ref{Ratio Fig}. Emission is coupled to excitation temperature, column density, and spatial extent, so without multiple transitions to constrain excitation temperatures, we cannot directly compare abundances between the two disks. With the exception of CS and N$_2$H$^+$ though, emission from the species we detect is likely optically thin, suggesting that integrated emission will be roughly proportional to column density, with some multiplicative offset due to the differing excitation conditions of the sources. We therefore use the flux density ratios as a rough proxy for abundance comparisons between the two sources.
        
        \subsubsection{$^{13}$C$^{18}$O, N$_2$H$^+$ $\&$ H$_2$CO}
        \label{SLS_S511}
            Strong $^{13}$C$^{18}$O emission is detected toward MWC~480 with none detected toward LkCa~15. Our limits on $^{13}$C$^{18}$O in LkCa~15, yielding a lower limit flux ratio of 5.3, are roughly consistent with the C$^{18}$O flux ratio of $\sim$6 in the observations used as template filters (see \S\ref{SLS_S31}). These results suggest a more massive gas-phase reservoir of CO in MWC~480. This is not surprising, given both that MWC~480 is a more massive disk and that CO is likely frozen out in much of the cold LkCa~15 disk, while it will remain in the gas phase in the warmer MWC~480 disk. CO freeze-out and processing on grains via sequential hydrogenation \citep[e.g.,][]{Qi_2013} can also explain the much higher fluxes of N$_2$H$^+$ and H$_2$CO in LkCa~15. These data are further analyzed in Qi et al. (in prep).
        
        \subsubsection{Deuterated species}
        \label{SLS_S512}
            We find that DCO$^{+}$ 4--3 emission is much brighter in MWC~480 than in LkCa~15. \cite{Huang_2017} find similar integrated flux densities between the two disks, however, for the 3--2 transition of DCO$^{+}$. Combining our results with those in \cite{Huang_2017} yields a DCO$^{+}$ 4--3/3--2 flux ratio of $\sim$1.9$\pm0.3$ in MWC~480 and $\sim$1.25$\pm0.3$ in LkCa~15. This may suggest that the DCO$^{+}$ in MWC~480 is warmer, which would additionally be consistent with the broad distribution of DCO$^{+}$ in LkCa~15 and the more centrally peaked DCO$^{+}$ in MWC~480 (i.e. there is more DCO$^{+}$ present in the colder outer regions of LkCa 15). The presence of both warm and cold pathways contributing to DCO$^+$ abundances has also recently been demonstrated in \cite{Carney_2018} for the Herbig Ae disk system HD 169142.
            
            Similar to \cite{Huang_2017}, we also find that the MWC~480 / LkCa~15 flux density ratio of DCO$^{+}$ is much higher than that of DCN. \cite{Huang_2017} interpreted these results as possible evidence that both warm and cold pathways \citep{Roueff_2013} contribute significantly to the formation of DCN. The present detection DNC may offer a possible path for investigating this chemistry further. In the gas phase, DNC forms via reaction of H$_2$D$^+$ with HCN, while DCN forms via reaction of H$_2$D$^+$ with HNC \citep[e.g.][]{Willacy_2009}. Now that HCN, HNC, DCN, and DNC have all been detected in disks (and three of the four in LkCa~15), future targeted studies may be able to better constrain the chemistry of this small deuteration network.
        
        \subsubsection{S-bearing species}
        \label{SLS_S513}
            The main CS isotopologue is strongly detected toward both disks and is possibly optically thick at smaller radii \citep[e.g.,][]{Teague_2017, Liu_2018}, making it difficult to make an abundance comparison from this transition alone. A higher CS column density in LkCa~15 is supported, however, by the fact that the optically thin CS isotopologues $^{13}$CS and C$^{34}$S are only detected there. Emission from the main isotopologue is compact with small emission holes seen for both disks. Observations of CS 5--4 emission toward the edge-on Flying Saucer disk were reported by \cite{Dutrey_2017} to be vertically unresolved, suggestive of dominant emission from dense gas near the midplane. If this vertical distribution is also present in MWC~480, the central depression in the CS emission could be explained by continuum subtraction effects in the inner disk where the dust optical depths are high \citep[e.g.,][]{Cleeves_2016, Teague_2017}. 
    
            Interestingly, H$_2$CS is detected toward MWC~480 with a compact emission profile, even though we suspect this disk has a lower CS abundance. Unlike its analog H$_2$CO, H$_2$CS formation is thought to be dominated by the gas phase reaction of CH$_3$ with atomic S \citep{LeGal_2019}, so this discrepancy may be due to the warmer conditions and stronger radiation fields in MWC~480 enhancing the concentration of reactants. These observations are analyzed in more detail in the context of a larger survey of S-bearing species in \cite{LeGal_2019}.
            
            
        \subsubsection{Hydrocarbons and nitriles}
        \label{SLS_S514}
            The hydrocarbons c-C$_3$H$_2$ and C$_2$D as well as the nitriles HC$_3$N and CH$_3$CN are found to be brighter toward MWC~480. One possible cause of this would be an enhanced C/O ratio toward MWC~480, which would then result in increased hydrocarbon and nitrile production, as suggested by \cite{Du_2015}, \cite{Bergin_2016}, and \cite{Cleeves_2018}. We caution however that the upper state energies of the c-C$_3$H$_2$ and nitrile transitions are quite high (50--200 K), and thus the measured flux densities are likely to be biased by the warmer temperature of MWC~480. Both of these phenomena are discussed in greater detail in \cite{Bergner_2018}.

    \subsection{Double ringed structures in LkCa~15}
    \label{SLS_S52}
        N$_2$H$^+$, H$_2$CO, DCO$^+$, DCN, and DNC all show varying degrees of evidence for possible double ring structure in LkCa~15, with an inner ring near the edge of the dust cavity and a second ring around 180 AU, near the edge of the millimeter dust disk. This feature was previously reported by \cite{Huang_2017} for the 3--2 transition of DCN around LkCa~15, as well as transitions of other species around HD 163296 and IM Lup \citep[where a double ring of DCO$^+$ was first identified by][]{Oberg_2015}. \cite{Oberg_2015} and \cite{Cleeves_2016} suggest that dust evolution may expose the outer disk to increase levels of irradiation, increasing temperatures and allowing CO to return to the gas phase. If this is the case, it may explain why the chemically linked and CO-sensitive species N$_2$H$^+$ and H$_2$CO have double ring profiles that follow each other closely.
    
    \subsection{Where are the COMs?}
    \label{SLS_S53}
        Predictions from chemical networks such as \cite{Walsh_2014} and \cite{Furuya_2014} suggest that a number of larger molecular species (e.g., CH$_3$CN, CH$_3$OH, HNCO, HCOOH, and CH$_3$CHO) should have been detectable in our survey (assuming excitation temperatures of 10-40K). The detection then of only CH$_3$CN in one disk is therefore quite interesting, and appears to be in agreement with mounting observational evidence for suppressed COM emission in disks \citep[e.g.,][]{Carney_2019}. Although there is strong evidence for abundant grain surface formation of COM precursors such as H$_2$CO in protoplanetary disks \citep{Loomis_2015, Carney_2017, Oberg_2017}, the sole detection of CH$_3$OH thus far \citep{Walsh_2016} found a column density up to an order of magnitude lower than expected from models \citep{Walsh_2014, Furuya_2014}.
        
\begin{figure*}[ht!]
        \centering
        \includegraphics[width=\textwidth]{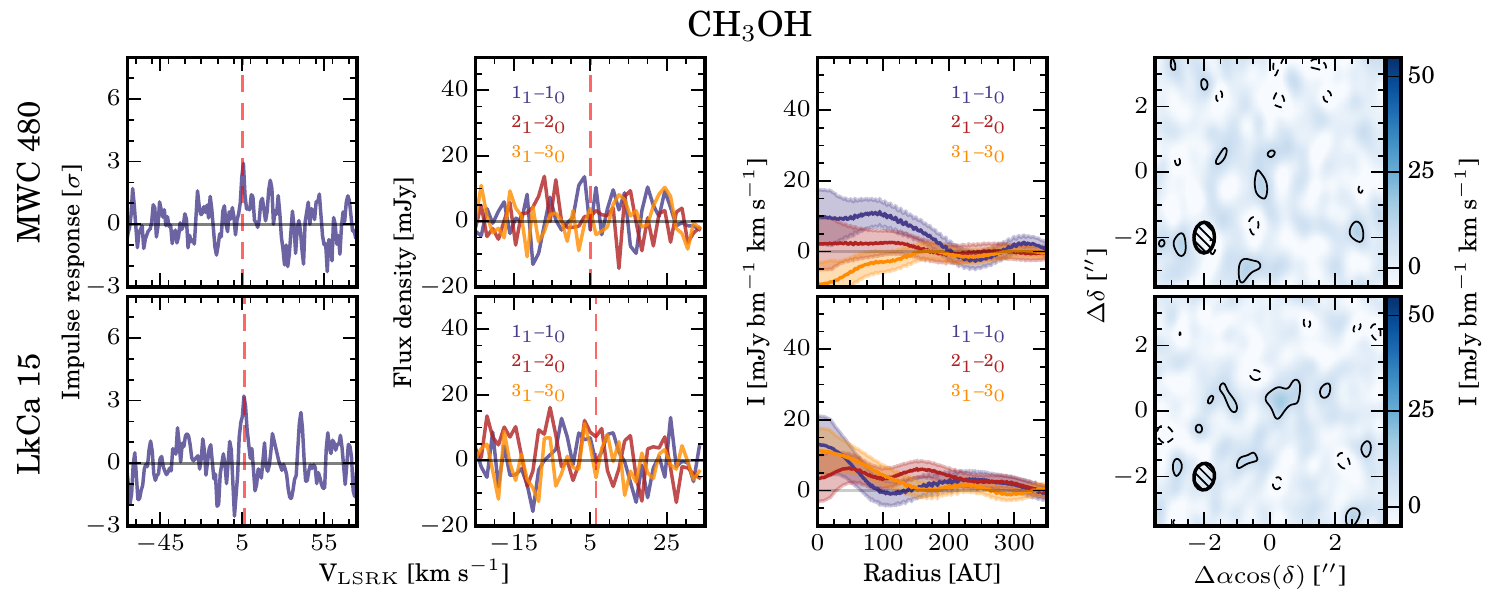}
        \caption{CH$_{3}$OH observations in MWC~480 and LkCa~15. Panel descriptions are identical to Figure \ref{13C18O Fig}, but with contours of $[-2,2,4,6,8,...]\times\sigma$.
        \label{CH3OH Fig}}
        \end{figure*}

        Searching for CH$_3$OH lines in our data yields no firm detections, as seen in Fig. \ref{CH3OH Fig}. Two tentative features are identified at $\sim$3$\sigma$ in the stacked spectra, but require further analysis and/or follow-up observations to be confirmed. Using the integrated flux uncertainties on each line (determined as detailed in \S3.3) and assuming an excitation temperature of 25~K\citep{Walsh_2016,Carney_2019}, we find a 3$\sigma$ upper limit column density of $\sim$8$\times$10$^{12}$ cm$^{-2}$, below the column densities expected from models \citep{Walsh_2014, Furuya_2014}. A more stringent analysis of these upper limits is presented within the context of a larger sample in Ilee et al. (in prep).
        
        These observations suggest that chemical models over-produce gas-phase abundances of O-bearing complex species. As recently discussed in \cite{Walsh_2017}, there are two plausible explanations. First, the rate coefficients assumed for O-bearing COM formation mechanisms on grain surfaces may be too high, due to e.g. overestimates of the internal radiation fields in disks. In this scenario, COMs may be present in the gas-phase, but at reduced abundances, mirroring their reduced ice abundances. The second possibility is that more complex species are not able to desorb off the grain surfaces in appreciable quantities. Recent studies suggests that larger molecules such as CH$_3$OH readily fragment when irradiated, and may not be able to efficiently photodesorb intact from grain surfaces \citep{Bertin_2016, Cruz-Diaz_2016}, an explanation which has been invoked to explain observed CH$_3$OH column densities toward TW Hya \citep{Walsh_2017, Ligterink_2018}. Future stacking analyses, deeper targeted integrations, or a combination of both are needed to obtain even limited COM inventories in disks, to explore COM/CH$_3$OH ratios, and CH$_3$OH and COM radial and vertical distributions. The latter will especially be key to distinguish between reduced-production and reduced-desorption scenarios, as the photodesorption efficiencies of COMs should depend on disk radius and height.

\section{Summary}
\label{SLS_S6}
    We have presented an overview of an unbiased interferometric spectral line survey toward the protoplanetary disks MWC~480 and LkCa~15. These results can be summarized as follows:
    
    \begin{enumerate}
    \item 24 transitions of 14 molecular species are detected, with 5 of these species (C$^{34}$S, $^{13}$CS, H$_{2}$CS, DNC, and C$_2$D) detected for the first time in a protoplanetary disk.
    \item Matched filtering of the survey using prior observations as templates allowed much faster line identification than a traditional imaging approach and improved SNR by factors of $\sim$2-3. 
    \item Significant differences are observed in the molecular inventories of MWC~480 and LkCa~15, which may be mostly explained by temperature differences between the disks.
    \item Species that require CO freeze-out are enhanced toward LkCa~15, while the detection of $^{13}$C$^{18}$O in MWC~480 suggests highly abundant gas-phase CO.
    \item S-bearing species are brighter toward LkCa~15, with the exception of H$_{2}$CS. Its presence in MWC~480 can likely be explained by warm gas-phase chemistry, which is investigated further in \cite{LeGal_2019}.
    \item Observed transitions of hydrocarbons and nitriles are brighter toward MWC~480. It is unclear if this is due to the high upper state energies of these transitions and therefore a temperature bias, or whether abundances are truly higher toward MWC~480, suggestive of an enhanced C/O ratio. Forward modeling (beyond the scope of this work) may be able to distinguish between these scenarios.
    \item Emission from COMs such as CH$_3$OH, CH$_3$CHO, and HCOOH is conspicuously absent, in conflict with column density predictions from chemical models, although two tentative features are found in the stacked CH$_3$OH spectra. These results are in line with the low CH$_3$OH column densities observed in TW Hya and discussed in \cite{Walsh_2017} and \cite{Carney_2019}, possibly suggesting that the desorption rates of O-bearing COMs in disk chemical models are currently overestimated. 
    \end{enumerate}

\acknowledgments
R.A.L. gratefully acknowledges funding from ALMA Student Observing Support. K.I.O. acknowledges funding from the David and Lucile Packard Foundation and from the Simons Foundation (SCOL \#321183). J.H. acknowledges support from the National Science Foundation Graduate Research Fellowship under Grant No. DGE-1144152. E.A.B. acknowledges funding through NSF grant AST-1514670 and NASA NNX16AB48G. C.W. acknowledges financial support from STFC (grant reference ST/R000549/1) and the University of Leeds. The National Radio Astronomy Observatory is a facility of the National Science Foundation operated under cooperative agreement by Associated Universities, Inc.  This paper makes use of the following ALMA data: ADS/JAO.ALMA\#2013.1.00226.S and ADS/JAO.ALMA\#2015.1.00657.S. ALMA is a partnership of ESO (representing its member states), NSF (USA) and NINS (Japan), together with NRC (Canada) and NSC and ASIAA (Taiwan), in cooperation with the Republic of Chile. The Joint ALMA Observatory is operated by ESO, AUI/NRAO and NAOJ.


\begin{thebibliography}{}
\expandafter\ifx\csname natexlab\endcsname\relax\def\natexlab#1{#1}\fi
\providecommand{\url}[1]{\href{#1}{#1}}

\end{thebibliography}


\begin{thebibliography}{}
\expandafter\ifx\csname natexlab\endcsname\relax\def\natexlab#1{#1}\fi
\providecommand{\url}[1]{\href{#1}{#1}}

\bibitem[{{Andrews} {et~al.}(2013){Andrews}, {Rosenfeld}, {Kraus}, \&
  {Wilner}}]{Andrews_2013}
{Andrews}, S.~M., {Rosenfeld}, K.~A., {Kraus}, A.~L., \& {Wilner}, D.~J. 2013,
  \apj, 771, 129

\bibitem[{{Belloche} {et~al.}(2008{\natexlab{a}}){Belloche}, {Comito},
  {Hieret}, {Menten}, {Mueller}, \& {Schilke}}]{Belloche_2008a}
{Belloche}, A., {Comito}, C., {Hieret}, C., {et~al.} 2008{\natexlab{a}}, ArXiv
  e-prints, arXiv:0801.3214

\bibitem[{{Belloche} {et~al.}(2009){Belloche}, {Garrod}, {M{\"u}ller},
  {Menten}, {Comito}, \& {Schilke}}]{Belloche_2009}
{Belloche}, A., {Garrod}, R.~T., {M{\"u}ller}, H.~S.~P., {et~al.} 2009, \aap,
  499, 215

\bibitem[{{Belloche} {et~al.}(2008{\natexlab{b}}){Belloche}, {Menten},
  {Comito}, {M{\"u}ller}, {Schilke}, {Ott}, {Thorwirth}, \&
  {Hieret}}]{Belloche_2008b}
{Belloche}, A., {Menten}, K.~M., {Comito}, C., {et~al.} 2008{\natexlab{b}},
  \aap, 482, 179

\bibitem[{{Belloche} {et~al.}(2017){Belloche}, {Meshcheryakov}, {Garrod},
  {Ilyushin}, {Alekseev}, {Motiyenko}, {Margul{\`e}s}, {M{\"u}ller}, \&
  {Menten}}]{Belloche_2017}
{Belloche}, A., {Meshcheryakov}, A.~A., {Garrod}, R.~T., {et~al.} 2017, \aap,
  601, A49

\bibitem[{{Bennett} \& {Kaiser}(2007)}]{Bennett_2007}
{Bennett}, C.~J., \& {Kaiser}, R.~I. 2007, \apj, 661, 899

\bibitem[{{Bergin} {et~al.}(2016){Bergin}, {Du}, {Cleeves}, {Blake}, {Schwarz},
  {Visser}, \& {Zhang}}]{Bergin_2016}
{Bergin}, E.~A., {Du}, F., {Cleeves}, L.~I., {et~al.} 2016, \apj, 831, 101

\bibitem[{{Bergner} {et~al.}(2018){Bergner}, {{Guzman}, V.~G. and {\"O}berg},
  {Loomis}, \& {Pegues}}]{Bergner_2018}
{Bergner}, J.~B., {{Guzman}, V.~G. and {\"O}berg}, K.~I., {Loomis}, R.~A., \&
  {Pegues}, J. 2018, \apj, In Press

\bibitem[{{Bergner} {et~al.}(2019){Bergner}, {{\"O}berg}, {Bergin}, {Loomis},
  {Pegues}, \& {Qi}}]{Bergner_2019}
{Bergner}, J.~B., {{\"O}berg}, K.~I., {Bergin}, E.~A., {et~al.} 2019, \apj,
  876, 25

\bibitem[{{Bertin} {et~al.}(2016){Bertin}, {Romanzin}, {Doronin}, {Philippe},
  {Jeseck}, {Ligterink}, {Linnartz}, {Michaut}, \& {Fillion}}]{Bertin_2016}
{Bertin}, M., {Romanzin}, C., {Doronin}, M., {et~al.} 2016, \apjl, 817, L12

\bibitem[{{Blake} {et~al.}(1986){Blake}, {Sutton}, {Masson}, \&
  {Phillips}}]{Blake_1986}
{Blake}, G.~A., {Sutton}, E.~C., {Masson}, C.~R., \& {Phillips}, T.~G. 1986,
  \apjs, 60, 357

\bibitem[{{Blake} {et~al.}(1994){Blake}, {van Dishoeck}, {Jansen}, {Groesbeck},
  \& {Mundy}}]{Blake_1994}
{Blake}, G.~A., {van Dishoeck}, E.~F., {Jansen}, D.~J., {Groesbeck}, T.~D., \&
  {Mundy}, L.~G. 1994, \apj, 428, 680

\bibitem[{{Booth} {et~al.}(2018){Booth}, {Walsh}, {Kama}, {Loomis}, {Maud}, \&
  {Juh{\'a}sz}}]{Booth_2018}
{Booth}, A.~S., {Walsh}, C., {Kama}, M., {et~al.} 2018, \aap, 611, A16

\bibitem[{{Carney} {et~al.}(2017){Carney}, {Hogerheijde}, {Loomis}, {Salinas},
  {{\"O}berg}, {Qi}, \& {Wilner}}]{Carney_2017}
{Carney}, M.~T., {Hogerheijde}, M.~R., {Loomis}, R.~A., {et~al.} 2017, ArXiv
  e-prints, arXiv:1705.10188

\bibitem[{{Carney} {et~al.}(2018){Carney}, {Fedele}, {Hogerheijde}, {Favre},
  {Walsh}, {Bruderer}, {Miotello}, {Murillo}, {Klaassen}, {Henning}, \& {van
  Dishoeck}}]{Carney_2018}
{Carney}, M.~T., {Fedele}, D., {Hogerheijde}, M.~R., {et~al.} 2018, \aap, 614,
  A106

\bibitem[{{Carney} {et~al.}(2019){Carney}, {Hogerheijde}, {Guzm{\'a}n},
  {Walsh}, {{\"O}berg}, {Fayolle}, {Cleeves}, {Carpenter}, \&
  {Qi}}]{Carney_2019}
{Carney}, M.~T., {Hogerheijde}, M.~R., {Guzm{\'a}n}, V.~V., {et~al.} 2019,
  \aap, 623, A124

\bibitem[{{Chiang} {et~al.}(2001){Chiang}, {Joung}, {Creech-Eakman}, {Qi},
  {Kessler}, {Blake}, \& {van Dishoeck}}]{Chiang_2001}
{Chiang}, E.~I., {Joung}, M.~K., {Creech-Eakman}, M.~J., {et~al.} 2001, \apj,
  547, 1077

\bibitem[{{Cleeves} {et~al.}(2015){Cleeves}, {Bergin}, {Qi}, {Adams}, \&
  {{\"O}berg}}]{Cleeves_2015}
{Cleeves}, L.~I., {Bergin}, E.~A., {Qi}, C., {Adams}, F.~C., \& {{\"O}berg},
  K.~I. 2015, \apj, 799, 204

\bibitem[{{Cleeves} {et~al.}(2016){Cleeves}, {{\"O}berg}, {Wilner}, {Huang},
  {Loomis}, {Andrews}, \& {Czekala}}]{Cleeves_2016}
{Cleeves}, L.~I., {{\"O}berg}, K.~I., {Wilner}, D.~J., {et~al.} 2016, \apj,
  832, 110

\bibitem[{{Cleeves} {et~al.}(2018){Cleeves}, {{\"O}berg}, {Wilner}, {Huang},
  {Loomis}, {Andrews}, \& {Guzman}}]{Cleeves_2018}
---. 2018, \apj, 865, 155

\bibitem[{{Cruz-Diaz} {et~al.}(2016){Cruz-Diaz}, {Mart{\'{\i}}n-Dom{\'e}nech},
  {Mu{\~n}oz Caro}, \& {Chen}}]{Cruz-Diaz_2016}
{Cruz-Diaz}, G.~A., {Mart{\'{\i}}n-Dom{\'e}nech}, R., {Mu{\~n}oz Caro}, G.~M.,
  \& {Chen}, Y.-J. 2016, \aap, 592, A68

\bibitem[{{Czekala} {et~al.}(2015){Czekala}, {Andrews}, {Jensen}, {Stassun},
  {Torres}, \& {Wilner}}]{Czekala_2015}
{Czekala}, I., {Andrews}, S.~M., {Jensen}, E.~L.~N., {et~al.} 2015, \apj, 806,
  154

\bibitem[{{Du} {et~al.}(2015){Du}, {Bergin}, \& {Hogerheijde}}]{Du_2015}
{Du}, F., {Bergin}, E.~A., \& {Hogerheijde}, M.~R. 2015, \apjl, 807, L32

\bibitem[{{Dutrey} {et~al.}(2007){Dutrey}, {Henning}, {Guilloteau}, {Semenov},
  {Pi{\'e}tu}, {Schreyer}, {Bacmann}, {Launhardt}, {Pety}, \&
  {Gueth}}]{Dutrey_2007}
{Dutrey}, A., {Henning}, T., {Guilloteau}, S., {et~al.} 2007, \aap, 464, 615

\bibitem[{{Dutrey} {et~al.}(2017){Dutrey}, {Guilloteau}, {Pi{\'e}tu},
  {Chapillon}, {Wakelam}, {Di Folco}, {Stoecklin}, {Denis-Alpizar}, {Gorti},
  {Teague}, {Henning}, {Semenov}, \& {Grosso}}]{Dutrey_2017}
{Dutrey}, A., {Guilloteau}, S., {Pi{\'e}tu}, V., {et~al.} 2017, \aap, 607, A130

\bibitem[{{Favre} {et~al.}(2018){Favre}, {Fedele}, {Semenov}, {Parfenov},
  {Codella}, {Ceccarelli}, {Bergin}, {Chapillon}, {Testi}, {Hersant},
  {Lefloch}, {Fontani}, {Blake}, {Cleeves}, {Qi}, {Schwarz}, \&
  {Taquet}}]{Favre_2018}
{Favre}, C., {Fedele}, D., {Semenov}, D., {et~al.} 2018, \apjl, 862, L2

\bibitem[{{Flaherty} {et~al.}(2018){Flaherty}, {Hughes}, {Teague}, {Simon},
  {Andrews}, \& {Wilner}}]{Flaherty_2018}
{Flaherty}, K.~M., {Hughes}, A.~M., {Teague}, R., {et~al.} 2018, \apj, 856, 117

\bibitem[{{Fuchs} {et~al.}(2009){Fuchs}, {Cuppen}, {Ioppolo}, {Romanzin},
  {Bisschop}, {Andersson}, {van Dishoeck}, \& {Linnartz}}]{Fuchs_2009}
{Fuchs}, G.~W., {Cuppen}, H.~M., {Ioppolo}, S., {et~al.} 2009, \aap, 505, 629

\bibitem[{{Furuya} \& {Aikawa}(2014)}]{Furuya_2014}
{Furuya}, K., \& {Aikawa}, Y. 2014, \apj, 790, 97

\bibitem[{{Gaia Collaboration} {et~al.}(2018){Gaia Collaboration}, {Brown},
  {Vallenari}, {Prusti}, {de Bruijne}, {Babusiaux}, {Bailer-Jones}, {Biermann},
  {Evans}, {Eyer}, \& et~al.}]{GAIA}
{Gaia Collaboration}, {Brown}, A.~G.~A., {Vallenari}, A., {et~al.} 2018, \aap,
  616, A1

\bibitem[{{Garrod} {et~al.}(2008){Garrod}, {Weaver}, \& {Herbst}}]{Garrod_2008}
{Garrod}, R.~T., {Weaver}, S.~L.~W., \& {Herbst}, E. 2008, \apj, 682, 283

\bibitem[{{Graninger} {et~al.}(2015){Graninger}, {{\"O}berg}, {Qi}, \&
  {Kastner}}]{Graninger_2015}
{Graninger}, D., {{\"O}berg}, K.~I., {Qi}, C., \& {Kastner}, J. 2015, \apjl,
  807, L15

\bibitem[{{Guilloteau} {et~al.}(2014){Guilloteau}, {Simon}, {Pi{\'e}tu}, {Di
  Folco}, {Dutrey}, {Prato}, \& {Chapillon}}]{Guilloteau_2014}
{Guilloteau}, S., {Simon}, M., {Pi{\'e}tu}, V., {et~al.} 2014, \aap, 567, A117

\bibitem[{{Guzm{\'a}n} {et~al.}(2017){Guzm{\'a}n}, {{\"O}berg}, {Huang},
  {Loomis}, \& {Qi}}]{Guzman_2017}
{Guzm{\'a}n}, V.~V., {{\"O}berg}, K.~I., {Huang}, J., {Loomis}, R., \& {Qi}, C.
  2017, \apj, 836, 30

\bibitem[{{Helling} {et~al.}(2014){Helling}, {Woitke}, {Rimmer}, {Kamp}, {Thi},
  \& {Meijerink}}]{Helling_2014}
{Helling}, C., {Woitke}, P., {Rimmer}, P.~B., {et~al.} 2014, Life, 4,
  arXiv:1403.4420

\bibitem[{{Herbig} \& {Bell}(1988)}]{Herbig_1988}
{Herbig}, G.~H., \& {Bell}, K.~R. 1988, {Third Catalog of Emission-Line Stars
  of the Orion Population : 3 : 1988}

\bibitem[{{Huang} {et~al.}(2017){Huang}, {{\"O}berg}, {Qi}, {Aikawa},
  {Andrews}, {Furuya}, {Guzm{\'a}n}, {Loomis}, {van Dishoeck}, \&
  {Wilner}}]{Huang_2017}
{Huang}, J., {{\"O}berg}, K.~I., {Qi}, C., {et~al.} 2017, \apj, 835, 231

\bibitem[{{Isella} {et~al.}(2012){Isella}, {P{\'e}rez}, \&
  {Carpenter}}]{Isella_2012}
{Isella}, A., {P{\'e}rez}, L.~M., \& {Carpenter}, J.~M. 2012, \apj, 747, 136

\bibitem[{{Jin} {et~al.}(2019){Jin}, {Isella}, {Huang}, {Li}, {Li}, \&
  {Ji}}]{Jin_2019}
{Jin}, S., {Isella}, A., {Huang}, P., {et~al.} 2019, \apj, 881, 108

\bibitem[{{Johansson} {et~al.}(1984){Johansson}, {Andersson}, {Ellder},
  {Friberg}, {Hjalmarson}, {Hoglund}, {Irvine}, {Olofsson}, \&
  {Rydbeck}}]{Johansson_1984}
{Johansson}, L.~E.~B., {Andersson}, C., {Ellder}, J., {et~al.} 1984, \aap, 130,
  227

\bibitem[{{J{\o}rgensen} {et~al.}(2016){J{\o}rgensen}, {van der Wiel},
  {Coutens}, {Lykke}, {M{\"u}ller}, {van Dishoeck}, {Calcutt}, {Bjerkeli},
  {Bourke}, {Drozdovskaya}, {Favre}, {Fayolle}, {Garrod}, {Jacobsen},
  {{\"O}berg}, {Persson}, \& {Wampfler}}]{Jorgensen_2016}
{J{\o}rgensen}, J.~K., {van der Wiel}, M.~H.~D., {Coutens}, A., {et~al.} 2016,
  \aap, 595, A117

\bibitem[{{Kaifu} {et~al.}(2004){Kaifu}, {Ohishi}, {Kawaguchi}, {Saito},
  {Yamamoto}, {Miyaji}, {Miyazawa}, {Ishikawa}, {Noumaru}, {Harasawa}, {Okuda},
  \& {Suzuki}}]{Kaifu_2004}
{Kaifu}, N., {Ohishi}, M., {Kawaguchi}, K., {et~al.} 2004, \pasj, 56, 69

\bibitem[{{Kastner} {et~al.}(2014){Kastner}, {Hily-Blant}, {Rodriguez},
  {Punzi}, \& {Forveille}}]{Kastner_2014}
{Kastner}, J.~H., {Hily-Blant}, P., {Rodriguez}, D.~R., {Punzi}, K., \&
  {Forveille}, T. 2014, \apj, 793, 55

\bibitem[{{Kastner} {et~al.}(2015){Kastner}, {Qi}, {Gorti}, {Hily-Blant},
  {Oberg}, {Forveille}, {Andrews}, \& {Wilner}}]{Kastner_2015}
{Kastner}, J.~H., {Qi}, C., {Gorti}, U., {et~al.} 2015, \apj, 806, 75

\bibitem[{{Kastner} {et~al.}(2018){Kastner}, {Qi}, {Dickson-Vandervelde},
  {Hily-Blant}, {Forveille}, {Andrews}, {Gorti}, {{\"O}berg}, \&
  {Wilner}}]{Kastner_2018}
{Kastner}, J.~H., {Qi}, C., {Dickson-Vandervelde}, D.~A., {et~al.} 2018, \apj,
  863, 106

\bibitem[{{Le Gal} {et~al.}(2019){Le Gal}, {{\"O}berg}, {Loomis}, {Pegues}, \&
  {Bergner}}]{LeGal_2019}
{Le Gal}, R., {{\"O}berg}, K.~I., {Loomis}, R.~A., {Pegues}, J., \& {Bergner},
  J.~B. 2019, \apj, 876, 72

\bibitem[{{Ligterink} {et~al.}(2018){Ligterink}, {Walsh}, {Bhuin},
  {Vissapragada}, {Terwisscha van Scheltinga}, \& {Linnartz}}]{Ligterink_2018}
{Ligterink}, N.~F.~W., {Walsh}, C., {Bhuin}, R.~G., {et~al.} 2018, \aap, 612,
  A88

\bibitem[{{Liu} {et~al.}(2018){Liu}, {Dipierro}, {Ragusa}, {Lodato}, {Herczeg},
  {Long}, {Harsono}, {Boehler}, {Menard}, {Johnstone}, {Pascucci}, {Pinilla},
  {Salyk}, {van der Plas}, {Cabrit}, {Fischer}, {Hendler}, {Manara}, {Nisini},
  {Rigliaco}, {Avenhaus}, {Banzatti}, \& {Gully-Santiago}}]{Liu_2018}
{Liu}, Y., {Dipierro}, G., {Ragusa}, E., {et~al.} 2018, arXiv e-prints,
  arXiv:1811.04074

\bibitem[{{Loomis} {et~al.}(2018{\natexlab{a}}){Loomis}, {\"Oberg}, {Andrews},
  {Walsh}, {Czekala}, {Huang}, \& {Rosenfeld}}]{VISIBLE_ASCL}
{Loomis}, R., {\"Oberg}, K., {Andrews}, S., {et~al.} 2018{\natexlab{a}},
  {VISIBLE: VISIbility Based Line Extraction}, , , ascl:1802.006

\bibitem[{{Loomis} {et~al.}(2018{\natexlab{b}}){Loomis}, {Cleeves},
  {{\"O}berg}, {Aikawa}, {Bergner}, {Furuya}, {Guzman}, \&
  {Walsh}}]{Loomis_2018_CH3CN}
{Loomis}, R.~A., {Cleeves}, L.~I., {{\"O}berg}, K.~I., {et~al.}
  2018{\natexlab{b}}, \apj, 859, 131

\bibitem[{{Loomis} {et~al.}(2015){Loomis}, {Cleeves}, {{\"O}berg}, {Guzman}, \&
  {Andrews}}]{Loomis_2015}
{Loomis}, R.~A., {Cleeves}, L.~I., {{\"O}berg}, K.~I., {Guzman}, V.~V., \&
  {Andrews}, S.~M. 2015, \apjl, 809, L25

\bibitem[{{Loomis} {et~al.}(2018{\natexlab{c}}){Loomis}, {{\"O}berg},
  {Andrews}, {Walsh}, {Czekala}, {Huang}, \& {Rosenfeld}}]{Loomis_2018_MF}
{Loomis}, R.~A., {{\"O}berg}, K.~I., {Andrews}, S.~M., {et~al.}
  2018{\natexlab{c}}, ArXiv e-prints, arXiv:1803.04987

\bibitem[{{Loomis} {et~al.}(2013){Loomis}, {Zaleski}, {Steber}, {Neill},
  {Muckle}, {Harris}, {Hollis}, {Jewell}, {Lattanzi}, {Lovas}, {Martinez},
  {McCarthy}, {Remijan}, {Pate}, \& {Corby}}]{Loomis_2013}
{Loomis}, R.~A., {Zaleski}, D.~P., {Steber}, A.~L., {et~al.} 2013, \apjl, 765,
  L9

\bibitem[{{Luhman} {et~al.}(2010){Luhman}, {Allen}, {Espaillat}, {Hartmann}, \&
  {Calvet}}]{Luhman_2010}
{Luhman}, K.~L., {Allen}, P.~R., {Espaillat}, C., {Hartmann}, L., \& {Calvet},
  N. 2010, \apjs, 186, 111

\bibitem[{{Mannings} \& {Sargent}(1997)}]{Mannings_1997}
{Mannings}, V., \& {Sargent}, A.~I. 1997, \apj, 490, 792

\bibitem[{{McGuire}(2018)}]{McGuire_2018}
{McGuire}, B.~A. 2018, \apjs, 239, 17

\bibitem[{{McGuire} {et~al.}(2016){McGuire}, {Carroll}, {Loomis}, {Finneran},
  {Jewell}, {Remijan}, \& {Blake}}]{McGuire_2016}
{McGuire}, B.~A., {Carroll}, P.~B., {Loomis}, R.~A., {et~al.} 2016, Science,
  352, 1449

\bibitem[{{McGuire} {et~al.}(2012){McGuire}, {Loomis}, {Charness}, {Corby},
  {Blake}, {Hollis}, {Lovas}, {Jewell}, \& {Remijan}}]{McGuire_2012}
{McGuire}, B.~A., {Loomis}, R.~A., {Charness}, C.~M., {et~al.} 2012, \apjl,
  758, L33

\bibitem[{{Miotello} {et~al.}(2016){Miotello}, {van Dishoeck}, {Kama}, \&
  {Bruderer}}]{Miotello_2016}
{Miotello}, A., {van Dishoeck}, E.~F., {Kama}, M., \& {Bruderer}, S. 2016,
  \aap, 594, A85

\bibitem[{{Miotello} {et~al.}(2017){Miotello}, {van Dishoeck}, {Williams},
  {Ansdell}, {Guidi}, {Hogerheijde}, {Manara}, {Tazzari}, {Testi}, {van der
  Marel}, \& {van Terwisga}}]{Miotello_2017}
{Miotello}, A., {van Dishoeck}, E.~F., {Williams}, J.~P., {et~al.} 2017, \aap,
  599, A113

\bibitem[{North(1963)}]{North_1963}
North, D.~O. 1963, Proceedings of the IEEE, 51, 1016

\bibitem[{{{\"O}berg} {et~al.}(2009){{\"O}berg}, {Garrod}, {van Dishoeck}, \&
  {Linnartz}}]{Oberg_2009}
{{\"O}berg}, K.~I., {Garrod}, R.~T., {van Dishoeck}, E.~F., \& {Linnartz}, H.
  2009, \aap, 504, 891

\bibitem[{{{\"O}berg} {et~al.}(2015){{\"O}berg}, {Guzm{\'a}n}, {Furuya}, {Qi},
  {Aikawa}, {Andrews}, {Loomis}, \& {Wilner}}]{Oberg_2015}
{{\"O}berg}, K.~I., {Guzm{\'a}n}, V.~V., {Furuya}, K., {et~al.} 2015, \nat,
  520, 198

\bibitem[{{{\"O}berg} {et~al.}(2010){{\"O}berg}, {Qi}, {Fogel}, {Bergin},
  {Andrews}, {Espaillat}, {van Kempen}, {Wilner}, \& {Pascucci}}]{Oberg_2010}
{{\"O}berg}, K.~I., {Qi}, C., {Fogel}, J.~K.~J., {et~al.} 2010, \apj, 720, 480

\bibitem[{{{\"O}berg} {et~al.}(2011){{\"O}berg}, {Qi}, {Fogel}, {Bergin},
  {Andrews}, {Espaillat}, {Wilner}, {Pascucci}, \& {Kastner}}]{Oberg_2011}
---. 2011, \apj, 734, 98

\bibitem[{{{\"O}berg} {et~al.}(2017){{\"O}berg}, {Guzm{\'a}n}, {Merchantz},
  {Qi}, {Andrews}, {Cleeves}, {Huang}, {Loomis}, {Wilner}, {Brinch}, \&
  {Hogerheijde}}]{Oberg_2017}
{{\"O}berg}, K.~I., {Guzm{\'a}n}, V.~V., {Merchantz}, C.~J., {et~al.} 2017,
  \apj, 839, 43

\bibitem[{{Ossenkopf} {et~al.}(2010){Ossenkopf}, {M{\"u}ller}, {Lis},
  {Schilke}, {Bell}, {Bruderer}, {Bergin}, {Ceccarelli}, {Comito}, {Stutzki},
  {Bacman}, {Baudry}, {Benz}, {Benedettini}, {Berne}, {Blake}, {Boogert},
  {Bottinelli}, {Boulanger}, {Cabrit}, {Caselli}, {Caux}, {Cernicharo},
  {Codella}, {Coutens}, {Crimier}, {Crockett}, {Daniel}, {Demyk}, {Dieleman},
  {Dominik}, {Dubernet}, {Emprechtinger}, {Encrenaz}, {Falgarone}, {France},
  {Fuente}, {Gerin}, {Giesen}, {di Giorgio}, {Goicoechea}, {Goldsmith},
  {G{\"u}sten}, {Harris}, {Helmich}, {Herbst}, {Hily-Blant}, {Jacobs}, {Jacq},
  {Joblin}, {Johnstone}, {Kahane}, {Kama}, {Klein}, {Klotz}, {Kramer},
  {Langer}, {Lefloch}, {Leinz}, {Lorenzani}, {Lord}, {Maret}, {Martin},
  {Martin-Pintado}, {McCoey}, {Melchior}, {Melnick}, {Menten}, {Mookerjea},
  {Morris}, {Murphy}, {Neufeld}, {Nisini}, {Pacheco}, {Pagani}, {Parise},
  {Pearson}, {P{\'e}rault}, {Phillips}, {Plume}, {Quin}, {Rizzo}, {R{\"o}llig},
  {Salez}, {Saraceno}, {Schlemmer}, {Simon}, {Schuster}, {van der Tak},
  {Tielens}, {Teyssier}, {Trappe}, {Vastel}, {Viti}, {Wakelam}, {Walters},
  {Wang}, {Whyborn}, {van der Wiel}, {Yorke}, {Yu}, \&
  {Zmuidzinas}}]{Ossenkopf_2010}
{Ossenkopf}, V., {M{\"u}ller}, H.~S.~P., {Lis}, D.~C., {et~al.} 2010, \aap,
  518, L111

\bibitem[{{Pety} {et~al.}(2012){Pety}, {Gratier}, {Guzm{\'a}n}, {Roueff},
  {Gerin}, {Goicoechea}, {Bardeau}, {Sievers}, {Le Petit}, {Le Bourlot},
  {Belloche}, \& {Talbi}}]{Pety_2012}
{Pety}, J., {Gratier}, P., {Guzm{\'a}n}, V., {et~al.} 2012, \aap, 548, A68

\bibitem[{{Phuong} {et~al.}(2018){Phuong}, {Chapillon}, {Majumdar}, {Dutrey},
  {Guilloteau}, {Pi{\'e}tu}, {Wakelam}, {Diep}, {Tang}, {Beck}, \&
  {Bary}}]{Phuong_2018}
{Phuong}, N.~T., {Chapillon}, E., {Majumdar}, L., {et~al.} 2018, \aap, 616, L5

\bibitem[{{Pi{\'e}tu} {et~al.}(2007){Pi{\'e}tu}, {Dutrey}, \&
  {Guilloteau}}]{Pietu_2007}
{Pi{\'e}tu}, V., {Dutrey}, A., \& {Guilloteau}, S. 2007, \aap, 467, 163

\bibitem[{{Pinte} {et~al.}(2018){Pinte}, {M{\'e}nard}, {Duch{\^e}ne}, {Hill},
  {Dent}, {Woitke}, {Maret}, {van der Plas}, {Hales}, {Kamp}, {Thi}, {de
  Gregorio-Monsalvo}, {Rab}, {Quanz}, {Avenhaus}, {Carmona}, \&
  {Casassus}}]{Pinte_2018}
{Pinte}, C., {M{\'e}nard}, F., {Duch{\^e}ne}, G., {et~al.} 2018, \aap, 609, A47

\bibitem[{{Punzi} {et~al.}(2015){Punzi}, {Hily-Blant}, {Kastner}, {Sacco}, \&
  {Forveille}}]{Punzi_2015}
{Punzi}, K.~M., {Hily-Blant}, P., {Kastner}, J.~H., {Sacco}, G.~G., \&
  {Forveille}, T. 2015, \apj, 805, 147

\bibitem[{{Qi} {et~al.}(2015){Qi}, {{\"O}berg}, {Andrews}, {Wilner}, {Bergin},
  {Hughes}, {Hogherheijde}, \& {D'Alessio}}]{Qi_2015}
{Qi}, C., {{\"O}berg}, K.~I., {Andrews}, S.~M., {et~al.} 2015, \apj, 813, 128

\bibitem[{{Qi} {et~al.}(2013{\natexlab{a}}){Qi}, {{\"O}berg}, \&
  {Wilner}}]{Qi_2013_H2CO}
{Qi}, C., {{\"O}berg}, K.~I., \& {Wilner}, D.~J. 2013{\natexlab{a}}, \apj, 765,
  34

\bibitem[{{Qi} {et~al.}(2013{\natexlab{b}}){Qi}, {{\"O}berg}, \&
  {Wilner}}]{Qi_2013}
---. 2013{\natexlab{b}}, \apj, 765, 34

\bibitem[{{Qi} {et~al.}(2013{\natexlab{c}}){Qi}, {{\"O}berg}, {Wilner}, \&
  {Rosenfeld}}]{Qi_2013_C3H2}
{Qi}, C., {{\"O}berg}, K.~I., {Wilner}, D.~J., \& {Rosenfeld}, K.~A.
  2013{\natexlab{c}}, \apjl, 765, L14

\bibitem[{{Qi} {et~al.}(2013{\natexlab{d}}){Qi}, {{\"O}berg}, {Wilner},
  {D'Alessio}, {Bergin}, {Andrews}, {Blake}, {Hogerheijde}, \& {van
  Dishoeck}}]{Qi_2013_Sci}
{Qi}, C., {{\"O}berg}, K.~I., {Wilner}, D.~J., {et~al.} 2013{\natexlab{d}},
  Science, 341, 630

\bibitem[{{Remijan} {et~al.}(2009){Remijan}, {Hollis}, {Jewell}, \&
  {Lovas}}]{Remijan_2009}
{Remijan}, A.~J., {Hollis}, J.~M., {Jewell}, P.~R., \& {Lovas}, F.~J. 2009, in
  64th International Symposium On Molecular Spectroscopy, RG04

\bibitem[{{Rosenfeld} {et~al.}(2013){Rosenfeld}, {Andrews}, {Hughes}, {Wilner},
  \& {Qi}}]{Rosenfeld_2013}
{Rosenfeld}, K.~A., {Andrews}, S.~M., {Hughes}, A.~M., {Wilner}, D.~J., \&
  {Qi}, C. 2013, \apj, 774, 16

\bibitem[{{Rosenfeld} {et~al.}(2012){Rosenfeld}, {Qi}, {Andrews}, {Wilner},
  {Corder}, {Dullemond}, {Lin}, {Hughes}, {D'Alessio}, \&
  {Ho}}]{Rosenfeld_2012}
{Rosenfeld}, K.~A., {Qi}, C., {Andrews}, S.~M., {et~al.} 2012, \apj, 757, 129

\bibitem[{{Roueff} {et~al.}(2013){Roueff}, {Gerin}, {Lis}, {Wootten},
  {Marcelino}, {Cernicharo}, \& {Tercero}}]{Roueff_2013}
{Roueff}, E., {Gerin}, M., {Lis}, D.~C., {et~al.} 2013, Journal of Physical
  Chemistry A, 117, 9959

\bibitem[{{Schwarz} {et~al.}(2018){Schwarz}, {Bergin}, {Cleeves}, {Zhang},
  {{\"O}berg}, {Blake}, \& {Anderson}}]{Schwarz_2018}
{Schwarz}, K.~R., {Bergin}, E.~A., {Cleeves}, L.~I., {et~al.} 2018, \apj, 856,
  85

\bibitem[{{Simon} {et~al.}(2000){Simon}, {Dutrey}, \&
  {Guilloteau}}]{Simon_2000}
{Simon}, M., {Dutrey}, A., \& {Guilloteau}, S. 2000, \apj, 545, 1034

\bibitem[{{Teague} {et~al.}(2016){Teague}, {Guilloteau}, {Semenov}, {Henning},
  {Dutrey}, {Pi{\'e}tu}, {Birnstiel}, {Chapillon}, {Hollenbach}, \&
  {Gorti}}]{Teague_2016}
{Teague}, R., {Guilloteau}, S., {Semenov}, D., {et~al.} 2016, \aap, 592, A49

\bibitem[{{Teague} {et~al.}(2017){Teague}, {Semenov}, {Gorti}, {Guilloteau},
  {Henning}, {Birnstiel}, {Dutrey}, {van Boekel}, \& {Chapillon}}]{Teague_2017}
{Teague}, R., {Semenov}, D., {Gorti}, U., {et~al.} 2017, \apj, 835, 228

\bibitem[{{The} {et~al.}(1994){The}, {de Winter}, \& {Perez}}]{The_1994}
{The}, P.~S., {de Winter}, D., \& {Perez}, M.~R. 1994, \aaps, 104, 315

\bibitem[{{Thi} {et~al.}(2004){Thi}, {van Zadelhoff}, \& {van
  Dishoeck}}]{Thi_2004}
{Thi}, W.-F., {van Zadelhoff}, G.-J., \& {van Dishoeck}, E.~F. 2004, \aap, 425,
  955

\bibitem[{{van Dishoeck} {et~al.}(1995){van Dishoeck}, {Blake}, {Jansen}, \&
  {Groesbeck}}]{vanDishoeck_1995}
{van Dishoeck}, E.~F., {Blake}, G.~A., {Jansen}, D.~J., \& {Groesbeck}, T.~D.
  1995, \apj, 447, 760

\bibitem[{{van't Hoff} {et~al.}(2017){van't Hoff}, {Walsh}, {Kama}, {Facchini},
  \& {van Dishoeck}}]{vant_Hoff_2017}
{van't Hoff}, M.~L.~R., {Walsh}, C., {Kama}, M., {Facchini}, S., \& {van
  Dishoeck}, E.~F. 2017, \aap, 599, A101

\bibitem[{{Walsh} {et~al.}(2016){Walsh}, {Juh{\'a}sz}, {Meeus}, {Dent}, {Maud},
  {Aikawa}, {Millar}, \& {Nomura}}]{Walsh_2016}
{Walsh}, C., {Juh{\'a}sz}, A., {Meeus}, G., {et~al.} 2016, ArXiv e-prints,
  arXiv:1609.02011

\bibitem[{{Walsh} {et~al.}(2014){Walsh}, {Millar}, {Nomura}, {Herbst}, {Widicus
  Weaver}, {Aikawa}, {Laas}, \& {Vasyunin}}]{Walsh_2014}
{Walsh}, C., {Millar}, T.~J., {Nomura}, H., {et~al.} 2014, \aap, 563, A33

\bibitem[{{Walsh} {et~al.}(2017){Walsh}, {Vissapragada}, \&
  {McGee}}]{Walsh_2017}
{Walsh}, C., {Vissapragada}, S., \& {McGee}, H. 2017, ArXiv e-prints,
  arXiv:1710.01219

\bibitem[{{Watanabe} \& {Kouchi}(2002)}]{Watanabe_2002}
{Watanabe}, N., \& {Kouchi}, A. 2002, \apjl, 571, L173

\bibitem[{{Willacy} \& {Woods}(2009)}]{Willacy_2009}
{Willacy}, K., \& {Woods}, P.~M. 2009, \apj, 703, 479

\bibitem[{{Yu} {et~al.}(2017){Yu}, {Evans}, {Dodson-Robinson}, {Willacy}, \&
  {Turner}}]{Yu_2017}
{Yu}, M., {Evans}, II, N.~J., {Dodson-Robinson}, S.~E., {Willacy}, K., \&
  {Turner}, N.~J. 2017, \apj, 841, 39

\bibitem[{{Zaleski} {et~al.}(2013){Zaleski}, {Seifert}, {Steber}, {Muckle},
  {Loomis}, {Corby}, {Martinez}, {Crabtree}, {Jewell}, {Hollis}, {Lovas},
  {Vasquez}, {Nyiramahirwe}, {Sciortino}, {Johnson}, {McCarthy}, {Remijan}, \&
  {Pate}}]{Zaleski_2013}
{Zaleski}, D.~P., {Seifert}, N.~A., {Steber}, A.~L., {et~al.} 2013, \apjl, 765,
  L10

\bibitem[{{Zhang} {et~al.}(2017){Zhang}, {Bergin}, {Blake}, {Cleeves}, \&
  {Schwarz}}]{Zhang_2017}
{Zhang}, K., {Bergin}, E.~A., {Blake}, G.~A., {Cleeves}, L.~I., \& {Schwarz},
  K.~R. 2017, Nature Astronomy, 1, 0130

\end{thebibliography}
\end{document}